\documentclass[preprint]{aastex}
\newcommand       \be           {\begin{equation}}
\newcommand       \ee           {\end{equation}}
\newcommand       \Angstrom     {\,{\rm \AA}}          
\newcommand       \eV           {\,{\rm eV}\,}
\newcommand       \K            {\,{\rm K}}
\newcommand       \cm           {\,{\rm cm}}
\newcommand       \s            {\,{\rm s}}
\newcommand       \erg          {\,{\rm erg}}
\newcommand	  \g		{\,{\rm g}}

\newcommand       \omegaT	{\omega_{\rm T}}

\newcommand	  \yr		{\,{\rm yr}}
\newcommand       \nH           {n_{\rm H}}

\newcommand	  \Tgas		{T_{\rm gas}}

\newcommand       \Qabs	        {Q_{\rm abs}}

\newcommand       \gtsim        {\gtrsim}
\newcommand       \ltsim        {\lesssim}

\newcommand	  \xB		{{\bf \hat{x}_{\rm B}}}
\newcommand	  \yB		{{\bf \hat{y}_{\rm B}}}
\newcommand	  \zB		{{\bf \hat{z}_{\rm B}}}
\newcommand	  \xJ		{{\bf \hat{x}_{\rm J}}}
\newcommand	  \yJ		{{\bf \hat{y}_{\rm J}}}
\newcommand	  \zJ		{{\bf \hat{z}_{\rm J}}}
\newcommand	  \bomega	{\mbox{\boldmath$\omega$\unboldmath}}
\newcommand	  \bmu  	{\mbox{\boldmath$\mu$\unboldmath}}
\newcommand	  \hatxi  	{\mbox{\boldmath$\hat{\xi}$\unboldmath}}
\newcommand	  \hatphi  	{\mbox{\boldmath$\hat{\phi}$\unboldmath}}
\newcommand	  \bJ		{{\bf J}}
\newcommand	  \bB		{{\bf B}}
\newcommand	  \bGam		{{\bf \Gamma}}
\newcommand	  \bQ		{{\bf Q}}
\newcommand       \ehat         {{\bf \hat{e}}}
\newcommand       \ahat         {{\bf \hat{a}}}
\newcommand	  \hatJ		{{\bf \hat{J}}}
\newcommand	  \Qspec        {{\bar{\tilde{\bQ}}_{\Gamma}}}
\newcommand	  \Qq		{{\langle \tilde{\bQ}_{\Gamma} \rangle_{\pm}}}

\newcommand	  \Fphiavgpm	{{\langle F \rangle_\pm^\phi}}
\newcommand	  \Hphiavgpm	{{\langle H \rangle_\pm^\phi}}
\newcommand	  \Fphiavgplus	{{\langle F \rangle_+^\phi}}
\newcommand	  \Hphiavgplus	{{\langle H \rangle_+^\phi}}
\newcommand	  \Fphiavgminus	{{\langle F \rangle_-^\phi}}
\newcommand	  \Hphiavgminus	{{\langle H \rangle_-^\phi}}

\newcommand	  \Pf		{{P_{\rm f}}}

\shortauthors{Weingartner \& Draine}
\shorttitle{Radiative Torques on Interstellar Grains. III.}

\begin{document}

\title{Radiative Torques on Interstellar Grains.  III.  Dynamics with 
Thermal Relaxation}

\author{Joseph C. Weingartner}
\affil{CITA, 60 St. George Street, University of
Toronto, Toronto, ON M5S 3H8, Canada; weingart@cita.utoronto.ca}

\and

\author{B.T. Draine}
\affil{Princeton University Observatory, Peyton Hall,
        Princeton, NJ 08544, USA; draine@astro.princeton.edu}

\begin{abstract}

In the previous papers in this series, we found that radiative torques
can play a major role in the alignment of grains with the interstellar 
magnetic field.  Since the radiative torques can
drive the grains to suprathermal
rotational speeds, in previous work
we made the simplifying assumption that the grain principal
axis of greatest moment of inertia is always parallel to the grain angular 
momentum.  This enabled us to describe many of the features of the grain 
dynamics.  However, this assumption fails when the grains enter periods 
of thermal rotation, which occur naturally in the radiative torque alignment 
scenario.  In the present paper, we relax this assumption and explore the 
consequences for the grain dynamics.  
We develop a treatment to follow the grain dynamics including thermal 
fluctuations and ``thermal flipping'', and show results for one illustrative
example.  By comparing with a treatment without thermal fluctuations, we
see that inclusion of thermal fluctuations can lead to qualitative changes
in the grain dynamics.  In a future installment in this series, we will use 
the more complete dynamical treatment developed here to perform a 
systematic study of grain alignment by radiative torques.

\end{abstract}


\section{Introduction}
\label{sec:intro}

Polarization of starlight by the interstellar medium was discovered
serendipitously (Hall 1949; Hall \& Mikesell 1949; Hiltner 1949a, b).
It was immediately recognized that the polarization must be due
to selective extinction by dust grains, and that this required that
interstellar dust grains be both nonspherical and aligned.
The first papers seeking to account for the observed alignment appeared almost
immediately 
(Spitzer \& Schatzman 1949; 
Spitzer \& Tukey 1949; 
Davis \& Greenstein 1951;
Spitzer \& Tukey 1951),
but a satisfactory explanation for the grain alignment has
eluded theoretical understanding for many decades, despite
numerous investigations (see Lazarian 2002 for a recent review).

Davis \& Greenstein
(1951) suggested that the alignment results from paramagnetic dissipation.  
For this mechanism to act, the grain must be paramagnetic and it must rotate
through the static interstellar magnetic field.  
Paramagnetism appears likely for amorphous silicate grains, and plausible
for hydrogenated carbonaceous grains.
Indeed, given the substantial fraction of the overall grain mass
contributed by Fe, a significant ferromagnetic or ferrimagnetic fraction 
would not be impossible.

Davis \& Greenstein assumed that the grain rotation is excited by elastic
impacts with gas atoms.  In this case, the energy in rotation about any 
grain axis is $\sim \frac{1}{2}k\Tgas$, where $k$ is Boltzmann's 
constant and $\Tgas$ is the gas temperature; thus, this motion is called
``thermal rotation''.  The grain rotating through a static magnetic field is 
analogous to a 
magnetic resonance experiment, in which a static solid is exposed to a 
rotating field.  In both cases, the material tries to magnetize along the
field direction, but with a lag, and energy is dissipated into heat.
In the magnetic resonance experiment, the dissipated energy originates in 
the radiation field, whereas in the interstellar case, the grain's 
rotational energy is dissipated.

The statistical mechanics of a rigid paramagnetic grain was analyzed
by Jones \& Spitzer (1967) and Purcell \& Spitzer (1971).
If the gas temperature $\Tgas$ exceeds the dust temperature $T_d$ and if 
there were no disalignment due to random collisions with gas atoms,
then the Davis-Greenstein mechanism would drive a grain to tend to
spin about its principal
axis of greatest moment of inertia (hereafter $\ahat_1$), which would 
tend to be aligned with the local interstellar magnetic field $\bB$.  

However, for typical conditions in the diffuse ISM, the timescale for 
disalignment (due to collisions with gas atoms) is substantially shorter than
the Davis-Greenstein alignment timescale $\tau_{\rm DG}$.  Jones \& Spitzer
(1967) showed that if grains have superparamagnetic inclusions (e.g., domains
of pure Fe) then $\tau_{\rm DG}$ could be reduced by several orders of 
magnitude.  Mathis (1986) noted that, since larger grains would be more likely
to contain one or more such inclusions, this could explain the observation
that relatively large grains ($a \gtsim 0.1 \micron$) are well aligned while
smaller grains ($a \ltsim 0.05 \micron$) are not 
(see, e.g., Kim \& Martin 1995).\footnote{Throughout this paper, we 
characterize the grain size by the radius $a$ of a sphere of equal volume.}

Research on the rotational dynamics of interstellar grains discovered
important effects which had hitherto been overlooked.
Martin (1971) pointed out that a charged spinning grain
has a magnetic moment parallel or anti-parallel to the angular velocity 
$\bomega$, so that the grain precesses about $\bB$.  The precession rate
is fast enough that, even if the actual grain alignment mechanism did not
involve $\bB$, the observed polarization would be either parallel or 
perpendicular to $\bB$.  Dolginov \& Mytrophanov (1976) showed that the
Barnett effect 
(the tendency for a spinning body to become magnetized antiparallel to
$\bomega$) provides a much larger magnetic moment.

The problem of grain alignment therefore cleanly separates into two issues:
(1) the alignment of grain angular momentum $\bJ$ with $\bB$, and
(2) the alignment of the grain itself with its instantaneous angular
momentum $\bJ$.

Purcell (1979) realized that the Barnett effect would lead rapidly-rotating
grains to tend to spin about 
$\ahat_1$, its principal axis of largest moment of inertia.
This is the configuration for which 
a grain with constant angular momentum $\bJ$ has minimum
rotational kinetic energy.
If the grain starts in a different configuration, then generally the 
magnetization due to the Barnett effect
changes direction (in a periodic manner) in grain body 
coordinates; the resulting paramagnetic dissipation (which Purcell called 
``Barnett dissipation'' in this case) drives the grain to its 
state with minimum rotational energy.\footnote{
	Even in the absence of Barnett dissipation, viscoelastic 
	dissipation would align $\ahat_1$ with $\bJ$; 
	Purcell found that Barnett 
	dissipation is stronger.}  
Purcell found that the alignment of $\ahat_1$
with $\bJ$ occurs on a much shorter timescale than $\tau_{\rm DG}$ or the
gas-drag timescale, if the grain rotational kinetic energy $\gg kT_d$.
The Barnett dissipation is necessarily accompanied by thermal fluctuations
which act to disalign $\ahat_1$ from $\bJ$ (Lazarian \& Roberge 1997).
Roberge \& Lazarian (1999) have calculated the Davis-Greenstein
alignment of oblate spheroidal grains with this effect included.

Purcell (1975, 1979) also pointed out that grains are subject to 
systematic torques fixed in grain body coordinates, and that these torques
can spin the grain up to suprathermal rotational speeds.  A 
suprathermally-rotating 
grain is largely impervious to disalignment by random gas atom bombardment,
and is thus free to undergo alignment on the Davis-Greenstein timescale.  
However, the systematic torques discussed by Purcell result from processes
that occur at the grain surface (e.g., the formation of H$_2$ at special 
surface sites, with subsequent ejection), and thus depend sensitively on 
the details of the grain surface.  Changes in the surface (e.g., due to 
accretion of atoms from the gas) can change the magnitude and direction 
of the systematic torque.  As a result, the grain will sometimes be spun
down, when the torque is directed opposite $\bomega$, and the grain 
rotational kinetic energy can be reduced to $\sim kT_{\rm gas}$ 
for a period of time.  Such episodes are called ``crossovers''.\footnote{The
systematic torque need not be directed {\it exactly} antiparallel to
$\bJ$ to result in a crossover; any torque component perpendicular to 
$\bJ$ averages to zero during the grain rotation.}

Crossovers were first studied by Spitzer \& McGlynn (1979).  They found that,
in the absence of stochastic torques, the direction of $\bJ$ remains constant.
The magnitude of $\bJ$ decreases and then increases again; during this episode
the grain flips over.  With stochastic torques, the
scenario is largely the same, except that the direction of $\bJ$ after
the crossover is not identical to its initial direction.  Spitzer \& 
McGlynn concluded that a grain would become completely disaligned after 
passing through a small number of crossovers.  This implies that if 
crossovers occur frequently (on a timscale shorter than $\tau_{\rm DG}$),
then paramagnetic relaxation (without superparamagnetic inclusions) is
ineffective at aligning grains.

Lazarian \& Draine (1997) pointed out that thermal fluctuations in a grain
can play an important role in crossovers.  In the process of Barnett 
dissipation, rotational energy is transferred to the lattice vibrational
modes, which act as a thermal reservoir.  Energy can also move in the
opposite direction; a Barnett fluctuation is the spontaneous transfer of 
some amount of energy from the vibrational modes to the grain rotation (at
constant $\bJ$).  Since the rotational energy is a function of the angle
$\gamma$
between $\ahat_1$ and $\bJ$,\footnote{For an axisymmetric grain with 
specified angular momentum and rotational energy, the angle $\gamma$ is 
constant.  The torque-free motion of a grain with arbitrary shape is more 
complicated; see \S \ref{sec:torque-free}.} 
thermal fluctuations ensure that $\ahat_1$ 
is never perfectly aligned with $\bJ$, even when the grain is rotating 
suprathermally.
Lazarian \& Draine found that the resulting small amount of disalignment
during periods of suprathermal rotation {\it decreases} the degree of 
disalignment that occurs during a crossover.  Essentially, the initial 
disalignment limits the minimum value that $\bJ$ assumes during the    
crossover, and thus limits the effect that random gas atom impacts can have
on a grain.  Lazarian \& Draine concluded that ordinary paramagnetic 
dissipation can indeed align interstellar grains.

Lazarian \& Draine (1999a) identified another potentially important 
consequence of Barnett fluctuations, which they called ``thermal flipping''.
Barnett fluctuations cause the grain to undergo a random walk in the angle
$\gamma$, and if the fluctuations are strong enough then the grain can 
flip.\footnote{%
	For an axisymmetric grain, a flip occurs when 
	$(\pi/2-\gamma)$ changes sign.
	See \S \ref{sec:grain_flips} for a 
description of flips in the general case of arbitary grain shape.}
Once a grain that is entering a crossover has flipped, the component
of the systematic torque along $\bJ$ is parallel to $\bJ$, rather than 
anti-parallel to $\bJ$, and the grain is
now spun up rather than down.  If a thermal flip occurs before $J$ becomes
very small, then disalignment can largely be avoided.  In the Spitzer \& 
McGlynn (1979) crossover scenario, $J$ goes to zero during the grain's flip;  
in the Lazarian \& Draine (1999a) scenario, the grain can flip at an earlier
stage and bypass the disaligning, low-$J$ stage.\footnote{Of course, a 
flip can also occur as a result of the random torques due to gas atom 
impacts.  However, the timescale for such flips to occur is much longer than
the timescale for flips due to Barnett fluctuations.  Furthermore, such 
flips only occur during a disaligning, low-$J$ stage.}

Lazarian \& Draine (1999a) pointed out that if a grain can thermally flip 
once, then perhaps it can do so twice.  Indeed, if the Barnett fluctuations
are strong enough, then the grain could rapidly flip back and forth.  
If the systematic torque (e.g., due to H$_2$ formation) is fixed in
grain body coordinates, then the torque reverses in inertial coordinates
each time the grain flips.  With rapid flipping, the systematic torque
will time-average to zero, and the grain is prevented from being spun up to
suprathermal rotation!
Lazarian \& Draine refer to this condition as ``thermal trapping''.   They 
found that grains smaller than a critical size $a_c \sim 0.01$--$0.1 \micron$
become thermally trapped, which could explain the observation that relatively
small grains are not aligned.  

Thus far, all paramagnetic effects had been assumed to be due to electrons.
Lazarian \& Draine (1999b) found that for relatively low rotational speeds
(as encountered during crossovers), nuclear paramagnetism can be much more 
important in Barnett dissipation and fluctuations.  Taking this into account,
Lazarian \& Draine (1999b) revised their estimate of the critical trapping
grain size to $a_c \gtsim 1 \micron$.  
Grains smaller than $\sim 1\micron$ are then expected to not be
aligned, in conflict with the observation that grains with $a \gtsim 0.1 
\micron$ are well aligned.

A resolution of this problem can be sought in another line of development
in grain alignment theory.  
Harwit (1970a,b) pointed out that absorption and emission of photons
could lead to a random walk in the grain angular momentum, with a tendency for
the grain angular momentum to be parallel to the Galactic plane.
Dolginov (1972) observed that individual grains might have chirality, with
different absorption and
scattering cross sections for left- and right-handed circularly polarized
light, so that anisotropic starlight might exert a systematic torque on
a grain.

In the first paper
of this series (Draine \& Weingartner 1996, hereafter Paper I), we 
evaluated the radiative torque (due to the absorption and scattering of 
starlight) exerted on an irregularly shaped grain with the optical properties
of ``astronomical silicate'' (Draine \& Lee 1984).
We found that if the interstellar radiation field is anisotropic at a level
of $\sim 10 \%$,\footnote{Weingartner \& Draine (2001) have estimated that 
the 
visible/UV radiation in the solar neighborhood is $\sim 10 \%$ anisotropic.} 
then radiative torques can dominate H$_2$ formation torques
for grains with $a\gtsim0.1\micron$.
Radiative torques are not effective at spinning up 
$a\ltsim 0.05\micron$
grains in regions where H is neutral; the strong opacity of the gas 
beyond the Lyman limit denies the small grains short-wavelength radiation with 
which they can strongly couple.

Radiative torques have clear advantages over H$_2$ formation torques as 
an agent for driving long-lived
suprathermal spin-up in the Purcell alignment scenario.
First, radiative torques depend on the global grain geometry and the 
starlight anisotropy direction, which are expected to be 
stable for $\gtsim 10^7\yr$, longer than the timescale for grain resurfacing.
Thus, spindowns and crossovers might occur less frequently if 
radiative torques are responsible for suprathermal rotation.  Still, even
one crossover can be deadly if thermal trapping occurs.  Since radiative 
torques depend on the starlight anisotropy direction, which is fixed in space
rather than in grain body coordinates, the 
radiative torque vector is not fixed in body coordinates, and therefore
does not exactly reverse in inertial coordinates
each time the grain flips.  Thus, grains spun
up by radiative torques might be impervious to thermal trapping.

Radiative torques have even greater potential; in the second paper of this
series (Draine \& Weingartner 1997, hereafter Paper II), we showed that 
they can {\it directly} align grains.  In fact, we found that direct alignment
by radiative torques likely dominates paramagnetic dissipation for grains
with $a \gtsim 0.1 \micron$.  In order to simplify the equations 
describing the grain dynamics, we assumed in Paper II that 
(1) $\ahat_1$ is always parallel to $\bJ$.  We also assumed that 
(2) the timescale for $\bJ$ to precess
about $\bB$ is much shorter than the other timescales involved in the 
dynamics, so that we could immediately average over the precession in the 
dynamical equations.
Given these assumptions, the dynamical state of the grain
can be completely specified by the angular speed $\omega$ and the angle 
$\xi_{\rm II}$ between $\ahat_1$ and $\bB$.  We constructed ``trajectory 
maps'' for three irregularly shaped grains (with $a=0.2 \micron$ and 
astronomical silicate composition) showing the evolution of grains initially 
characterized by arbitrary combinations of $(\omega, \xi_{\rm II})$.  The 
trajectories often terminate on stable stationary points characterized
by a particular $(\omega, \xi_{\rm II})$; i.e., the radiative torques often
result in grain alignment.  The trajectories also often pass
through $\omega = 0$, suggesting that crossovers (of a sort) occur naturally 
in the radiative torque alignment scenario.  

However, because of the simplifying assumptions adopted in Paper II, we were 
unable to follow the dynamics through these low-$J$ stages.  First, we ignored 
the stochastic nature of the torques due to H$_2$ formation and collisions
with gas atoms.  This stochasticity can be safely ignored during periods of
suprathermal rotation but should be included when $\omega \rightarrow 0$.
Second, the assumption that $\ahat_1 \parallel \bJ$ breaks down when 
$J \rightarrow 0$.  We would expect this
simplification to be more severe, since thermal 
flipping might prevent $\omega$ from becoming so small that the first 
simplification fails badly.

Here we generalize the treatment of Paper II by relaxing the assumption 
that $\ahat_1 \parallel \bJ$.  In future papers, we will add a treatment of 
the stochastic nature of gas atom impacts and H$_2$ formation and we will
use the resulting formalism to conduct a systematic study of grain 
alignment by radiative torques.
The outline of the present paper is as follows:

In \S \ref{sec:grain_dynamics}, we describe the general dynamics of an
asymmetric grain and how it responds to external torques.  Most studies of
grain alignment adopt axisymmetric grain shapes for simplicity, but in this
case the high degree of symmetry suppresses the radiative torques.  Thus,
we treat the most general grain shape, with no degeneracy in the eigenvalues 
of the inertia tensor.  
We explain the concept of ``flip state'' for the grain, and show that
the grain can change flip states only when the rotational kinetic
energy passes through a critical value.  Because the tumbling motion of
the grain is very rapid, it is necessary and appropriate to average over
the torque-free motion before considering the effects of any external
torques.  The torque-free motion includes exchange of energy
between rotational and vibrational modes through the
phenomenon of ``Barnett dissipation''.

In \S \ref{sec:torques}, we discuss the various external torques that act on 
interstellar grains, due to the Barnett magnetic moment, gas and IR emission
drag, H$_2$ formation, paramagnetic dissipation, and starlight.  

The rapid exchange of energy between rotational and vibrational modes
implies that even when the grain has a fixed angular momentum $\bJ$,
it is important to average over the different possible values of the
grain kinetic energy; this thermal
averaging is discussed in \S\ref{sec:q_avg}.
In \S\ref{sec:eq_of_motion}, we obtain the equations for the time evolution of
the angular momentum $J$ and the angle $\xi$ between $\bJ$ and $\bB$.

As already noted, the grain dynamics involves the possibility that thermal
fluctuations will cause the grain to change from one ``flip state'' to the
other.
In \S\ref{sec:strategy_just_flips}, we develop an algorithm for
evolving the grain dynamics, including stochastic flipping.
We present an algorithm that can be applied both in the limit where
flipping occurs many times even for the shortest feasible computational
time step, as well as in the limit where flipping occurs very rarely.

In \S \ref{sec:results}, we present results of the dynamical evolution for
one particular case (i.e., for a given grain shape, composition, and size;
angle $\psi$ between the magnetic field and the radiation anisotropy 
direction; and radiation field spectrum).  For this case, the analysis of 
Paper II 
yields two stable stationary points and two crossover points.  Applying the 
more complete analysis developed here, we find a new stable stationary point 
with a thermal rotational speed and no crossovers.  All of the trajectories 
terminate on either the new stationary point or on one of the two stable 
stationary points with suprathermal rotation.  

In \S \ref{sec:summary}, we summarize the results of this paper and 
discuss the remaining issues that must be addressed to complete our 
understanding of grain alignment by radiative torques.  

Since we employ a large number of physical quantities, we provide
a glossary of notation (Appendix E) for easy reference.  

\section{\label{sec:grain_dynamics} Grain Dynamics}

\subsection{Grain Geometry}

We consider the irregular grain shape from Paper I (``shape 1'' in Paper II).  
The moment of 
inertia tensor has eigenvalues $I_1 \ge I_2 \ge I_3$, with principal axes
$\ahat_1$, $\ahat_2$, and $\ahat_3$.  We define dimensionless parameters
$\alpha_j$ by 
\be
I_j \equiv \alpha_j \frac{2}{5} \rho V a^2~~~,
\ee
where $\rho$ is the density and $V = 4 \pi a^3/3$ is the grain volume.  

\subsection{Coordinate Systems}

\begin{figure}
\epsscale{1.00}
\plotone{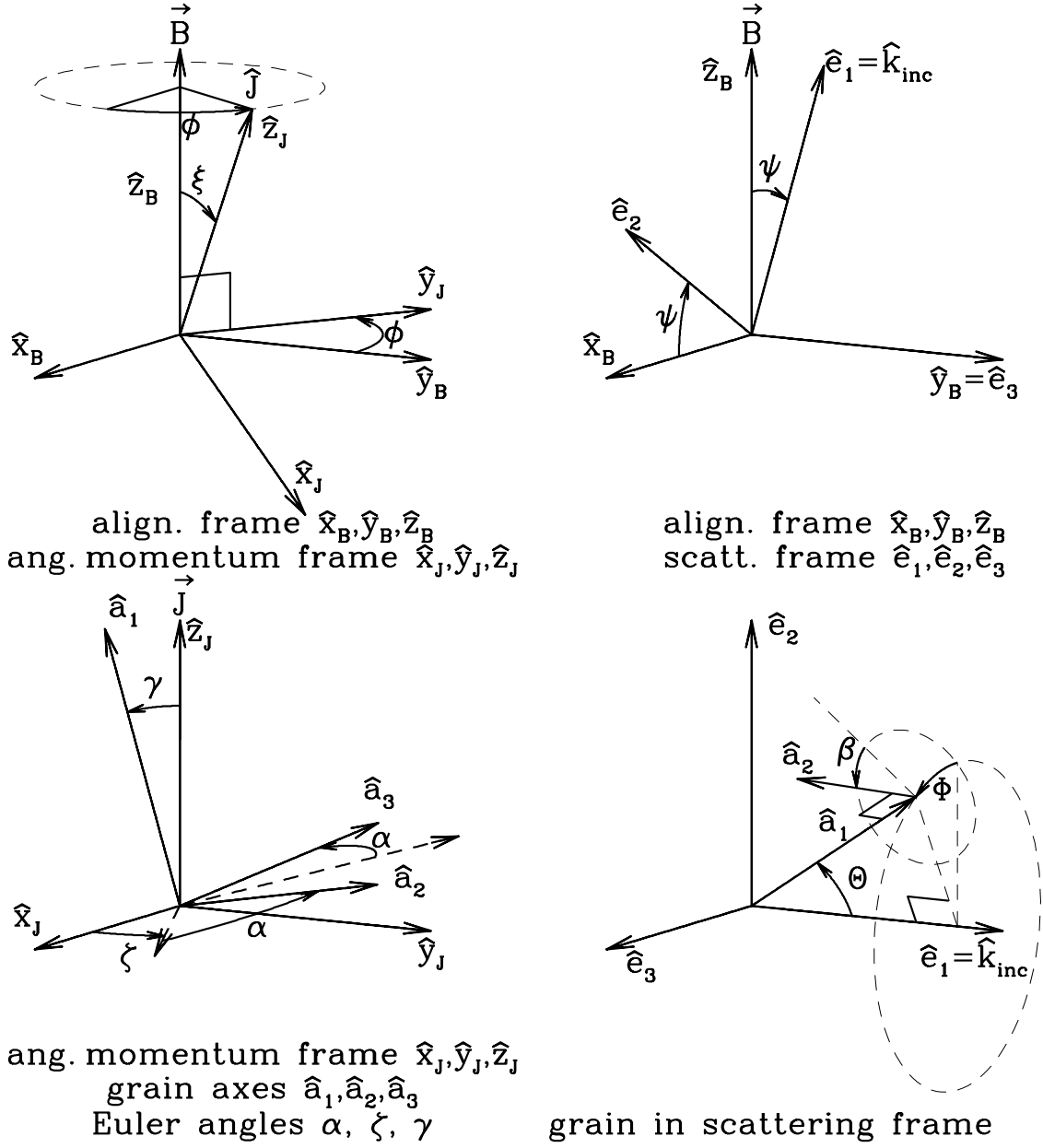}
\caption{
\label{fig:coord_frames}
Orientation of the ``angular momentum'' ($\xJ, \yJ, \zJ$)
and ``scattering'' ($\ehat_i$) coordinate frames
in the ``alignment'' ($\xB, \yB, \zB$) coordinate frame and orientation of
the grain axes $\ahat_i$ in the angular momentum frame and 
scattering frame.  
        }
\end{figure}

Since we are interested in grain alignment with the magnetic field $\bB$,
we adopt a coordinate system $x_{\rm B}$, $y_{\rm B}$, $z_{\rm B}$, with 
$\bB \parallel \zB$.  This system, which we call ``alignment''
coordinates, defines an inertial reference frame.  It will also be convenient
to define ``angular momentum'' coordinates $x_{\rm J}$, $y_{\rm J}$, 
$z_{\rm J}$, with $\zJ \parallel \bJ$.  We adopt the following 
transformation between these coordinate systems:
\be
\xB = \cos\xi \cos\phi \, \xJ - \sin\phi \, \yJ + \sin\xi \cos\phi \, \zJ~~~, 
\ee
\be
\yB = \cos\xi \sin\phi \, \xJ + \cos\phi \, \yJ + \sin\xi \sin\phi \, \zJ~~~,
\ee
\be
\zB = -\sin\xi \, \xJ + \cos\xi \, \zJ~~~,
\ee
where $\xi$ and $\phi$ are, respectively, the polar and azimuthal angles of
$\bJ$ in alignment coordinates.  Thus, we have the correspondence
$\xJ = \hatxi$, $\yJ = \hatphi$, $\zJ = \hatJ$.  See Figure 
\ref{fig:coord_frames}.

For computing radiative
torques, we define ``scattering'' coordinates $e_1$, $e_2$, $e_3$, with 
$\ehat_1$ parallel to the radiation propagation direction.  Three angles
are required to specify the grain orientation in this frame.  
Again, see Figure \ref{fig:coord_frames}.
The orientation
of $\ahat_1$ is described by the two angles $\Theta \in [0, \pi]$ and
$\Phi \in [0, 2\pi]$, where
\be
\ahat_1 = \cos\Theta \, \ehat_1 + \sin\Theta \cos\Phi \, \ehat_2 + 
\sin\Theta \sin\Phi \, \ehat_3~~~.
\ee
A third angle $\beta \in [0, 2\pi]$ describes rotation of $\ahat_2$ about
$\ahat_1$:
\be
\ahat_2 = - \sin\Theta \cos\beta \, \ehat_1 + (\cos \Theta \cos \Phi
\cos \beta - \sin \Phi \sin \beta) \, \ehat_2 + (\cos \Theta \sin \Phi 
\cos \beta + \cos \Phi \sin \beta) \, \ehat_3~~~.
\ee
Given $\ahat_i$ and $\ehat_i$, the angles $\Theta$, $\Phi$, and $\beta$ 
can be found as follows:
\be
\label{eq:Theta}
\Theta = \cos^{-1} (\ahat_1 \cdot \ehat_1)~~~,
\ee
\be
\label{eq:Phi}
\Phi = 2 \tan^{-1} \left( \frac{\sin \Theta - \ahat_1 \cdot \ehat_2}
{\ahat_1 \cdot \ehat_3} \right)~~~,
\ee
and
\be
\label{eq:beta}
\beta = 2 \tan^{-1} \left[ \frac{\sin \Theta + \ahat_2 \cdot 
\ehat_1 }{\sin \Theta \left( \ahat_2 \cdot \ehat_3 \cos \Phi
- \ahat_2 \cdot \ehat_2 \sin \Phi \right)} \right]~~~.
\ee
In Appendix A, we give expressions for $\Phi$ and $\beta$ when 
the denominators in equations (\ref{eq:Phi}) and (\ref{eq:beta}) are zero.

\subsection{Thermal Rotation}

In the absence of (a) internal dissipation and (b) any external torques 
besides those due to elastic collisions with gas atoms, the rms rotation rate 
for a sphere of radius $a$ is given by
\be
\label{eq:omega_T}
\omegaT = \left( \frac{15 k \Tgas}{8 \pi \rho a^5} \right)^{1/2} =
1.66 \times 10^5 \left(\frac{3 \g \cm^{-3}}{\rho}\right)^{1/2}
\left(\frac{\Tgas}{100\K}\right)^{1/2} \left(\frac{0.1 \micron}{a}\right)^{5/2}
\s^{-1}~~~.
\ee

\subsection{Dynamical Equations}

The equations describing the evolution of $\bJ$ and the rotational kinetic
energy $E$ with time are 
\be
\label{eq:dyn1}
\frac{d\bJ}{dt} = {\bf \Gamma}
\ee
and
\be
\label{eq:dyn2a}
\frac{dE}{dt} = \frac{1}{2} \left( \bJ \cdot \frac{d\bomega}{dt} + 
{\bf \Gamma} \cdot \bomega \right)~~~,
\ee
where ${\bf \Gamma}$ is the net torque on the grain.  Barnett dissipation 
and fluctuations are manifested in the $\bJ \cdot d\bomega/dt$ term.
It is convenient to introduce the following dimensionless quantity: 
\be
\label{eq:q_def}
q \equiv 2 I_1 E / J^2~~~.
\ee
Since the rotational energy lies between $J^2/2 I_1$ and $J^2/2 I_3$, 
$1 \le q \le I_1/I_3$.  The dynamical equation for $q$ is
\be
\label{eq:dyn2}
\frac{dq}{dt} = \frac{1}{J^2} \left[ I_1 \left( \bJ \cdot \frac{d\bomega}{dt}
+ \bGam \cdot \bomega \right) - 2 q \bJ \cdot \bGam \right]~~~.
\ee

Generally, ${\bomega}$ is not parallel to $\bJ$, 
in which case ${\bomega}$ executes some sort of periodic motion in grain  
body coordinates and the grain executes a more complicated quasi-periodic 
motion in space (see \S\S \ref{sec:euler_eqs} and \ref{sec:axis_motion}).
The timescale for this motion is usually much shorter than the 
timescale on which $\bJ$ changes due to the external torques (also 
$\omega^{-1}$ is usually much shorter than this latter timescale).  
Thus, we will average equations (\ref{eq:dyn1}) and (\ref{eq:dyn2}) over 
one ``cycle'' of torque-free motion.

We expect the terms $\bGam \cdot \bomega$ and $- 2 q \bJ \cdot \bGam /I_1$ 
in equation (\ref{eq:dyn2}) to be of the same order of magnitude, and to be 
small compared with the rate at which the rotational 
energy varies due to internal processes, so that only internal relaxation 
need be considered in determining the evolution of $q$.  (We use the term
``relaxation'' to refer to both dissipation and fluctuations.)  
We will examine the validity of this approximation on a torque-by-torque 
basis in \S \ref{sec:torques}.

\subsection{Torque-free Motion}
\label{sec:torque-free}

\subsubsection{\label{sec:euler_eqs} Euler Equations}

The motion of an asymmetric rigid body, when the net external torque is zero, 
is nicely described in \S 37 of Landau \& Lifshitz (1976).  First, the Euler 
equations are solved for the components of the angular velocity ${\bomega}$ in 
grain body coordinates.  The solution depends on the magnitude of the 
angular momentum ${\bf J}$ and the rotational kinetic energy $E$, and 
involves the Jacobi elliptic functions.  There are several different 
notational conventions for these functions in the literature; thus, we begin
by defining ours.  The elliptic integral of the first kind is given by
\be
F(\epsilon, m) \equiv \int_0^{\epsilon} d\theta \left( 1-m \sin^2 \theta 
\right)^{-1/2}~~~.
\ee
The Jacobi amplitude ${\rm am}(w, m)$ is defined as the inverse of the 
above integral; if $w = F(\epsilon, m)$, then $\epsilon = {\rm am}(w, m)$.  
The Jacobi elliptic functions are given by ${\rm sn}(w, m) = \sin\epsilon$,
${\rm cn}(w, m) = \cos\epsilon$, and ${\rm dn}(w, m) = 
(1-m \sin^2 \epsilon)^{1/2}$.

Next, we introduce the dimensionless quantity
\be
k^2 \equiv \frac{(I_2 - I_3) (q - 1)}{(I_1 - I_2) (1 - I_3 q /I_1)}~~~.
\ee

When $q < I_1/I_2$, $k^2 < 1$ and the solution of the Euler equations
is as follows:
\be
\label{eq:omega_1a}
\omega_1 = \pm \frac{J}{I_1} \left[ \frac{I_1 - I_3 q}{I_1 - I_3} \right]^{1/2}
{\rm dn}(\tau, k^2)~~~,
\ee
\be
\label{eq:omega_2a}
\omega_2 = - \frac{J}{I_2} \left[ \frac{I_2 (q-1)}{I_1 - I_2} \right]^{1/2}
{\rm sn}(\tau, k^2)~~~,
\ee
\be
\label{eq:omega_3a}
\omega_3 = (\pm)_1 \frac{J}{I_3} \left[ \frac{I_3 (q-1)}{I_1 - I_3} 
\right]^{1/2} {\rm cn}(\tau, k^2)~~~,
\ee
where
the $(\pm)_1$ in equation (\ref{eq:omega_3a}) stands for the same  
sign chosen in equation (\ref{eq:omega_1a}) and 
\be
\tau \equiv t J \left[ \frac{(I_1 - I_2) (1 -I_3 q/I_1)}
{I_1 I_2 I_3} \right]^{1/2}~~~.
\ee
The period in time $t$ is
\be
\label{eq:period1}
P_t = 4 F(\pi/2, k^2) \left[ \frac{I_1 I_2 I_3}{(I_1 - I_2) (1 -I_3 q/I_1)}
\right]^{1/2} J^{-1}
\ee
and the period in the variable $\tau$ is
\be 
P_{\tau} = 4 F(\pi/2, k^2)~~~.
\ee

When $q > I_1/I_2$,
\be
\label{eq:omega_1b}
\omega_1 = \pm \frac{J}{I_1} \left[ \frac{I_1 - I_3 q}{I_1 - I_3} \right]^{1/2}
{\rm cn}(\tau, k^{-2})~~~,
\ee
\be
\label{eq:omega_2b}
\omega_2 = - \frac{J}{I_2} \left[ \frac{I_2 (1-I_3 q/I_1)}
{I_2 - I_3} \right]^{1/2} {\rm sn}(\tau, k^{-2})~~~,
\ee
\be
\label{eq:omega_3b}
\omega_3 = (\pm)_1 \frac{J}{I_3} \left[ \frac{I_3 (q-1)}{I_1 - I_3} 
\right]^{1/2} {\rm dn}(\tau, k^{-2})~~~,
\ee
with
\be
\tau \equiv t J \left[ \frac{(I_2 - I_3) (q -1)}
{I_1 I_2 I_3} \right]^{1/2}~~~,
\ee
\be
\label{eq:period2}
P_t = 4 F(\pi/2, k^{-2}) \left[ \frac{I_1 I_2 I_3}{(I_2 - I_3) (q -1)}
\right]^{1/2} J^{-1}~~~,
\ee
and
\be
P_{\tau} = 4 F(\pi/2, k^{-2})~~~,
\ee
where $(\pm)_1$ in equation (\ref{eq:omega_3b}) stands for the sign choice in
equation (\ref{eq:omega_1b}).
Of course, when $q=I_1/I_2$, $\omega_1 = \omega_3 = 0$ and 
$\omega_2 = \pm J/I_2$.  

\begin{figure}
\epsscale{1.00}
\plotone{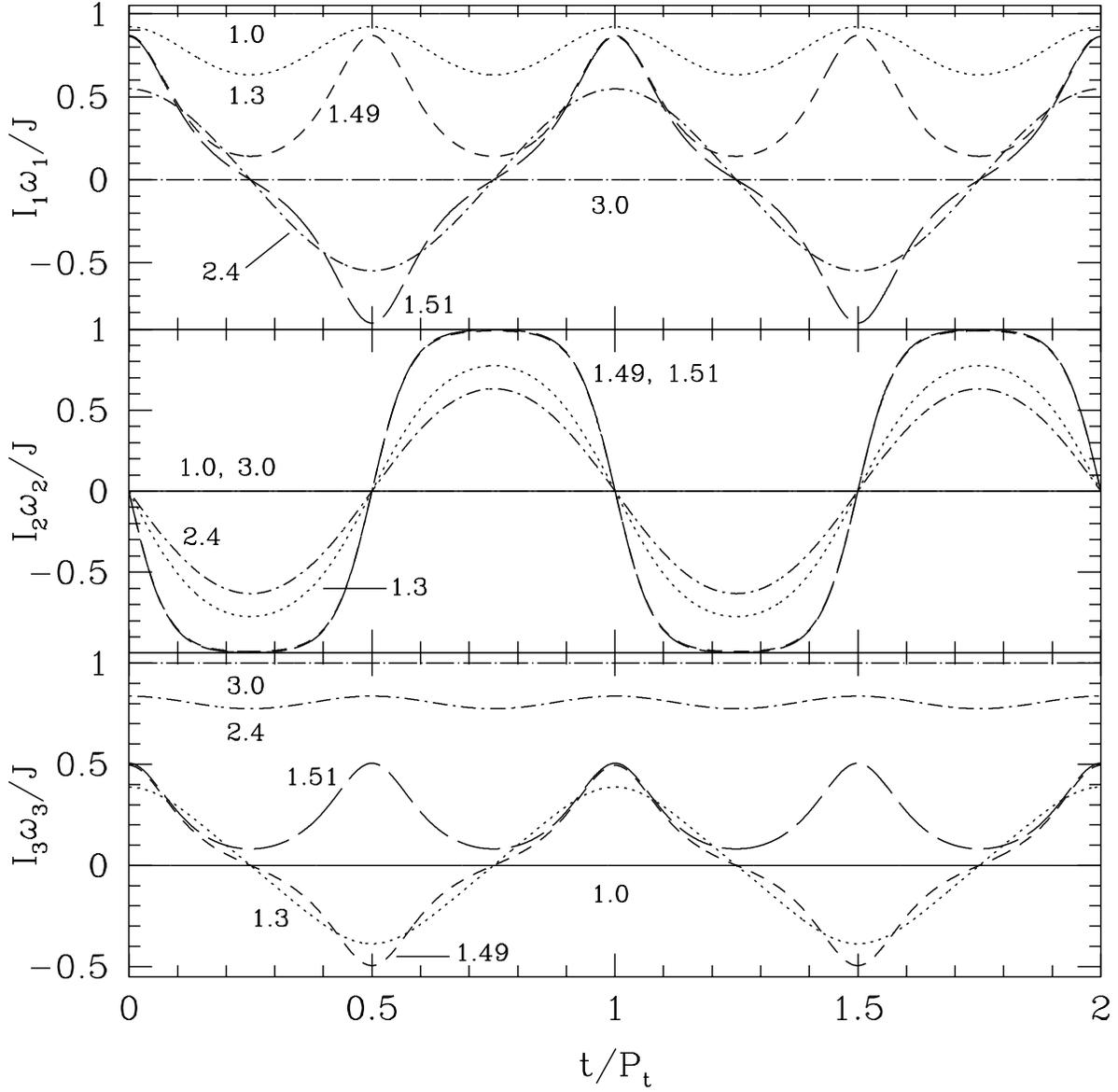}
\caption{
\label{fig:omega_example}
Components of the angular velocity along grain principal axes, for 
$I_1:I_2:I_3 = 3:2:1$.  Values of $q$ are indicated and positive flip
states are assumed.  
        }
\end{figure}

In Figure \ref{fig:omega_example}, we plot 
$\omega_i$ (with plus signs in eqs.~\ref{eq:omega_1a}, \ref{eq:omega_3a},
\ref{eq:omega_1b}, and \ref{eq:omega_3b})
over two periods for the case with $I_1:I_2:I_3 = 3:2:1$ and
several values of $q$.

\begin{figure}
\epsscale{1.00}
\plotone{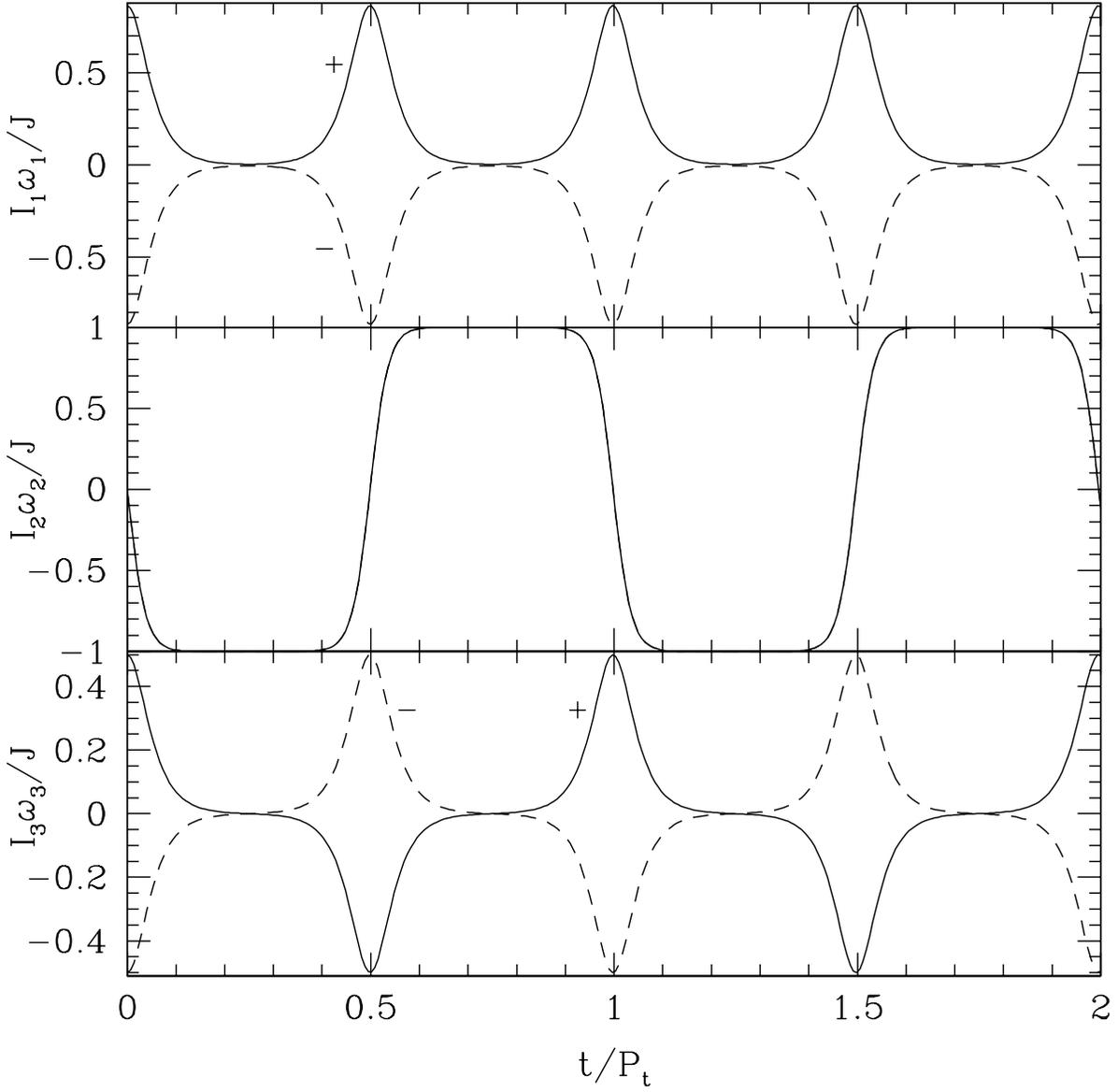}
\caption{
\label{fig:omega_flip_state}
Components of the angular velocity along grain principal axes, for
$I_1:I_2:I_3 = 3:2:1$ and $q$ slightly less than $I_1/I_2$; + (-) indicates 
the positive (negative) flip state with respect to $\ahat_1$.  $\omega_2$ 
does not depend on the flip state.
        }
\end{figure}

Note that when $q \rightarrow I_1/I_2$ from below [above], equations 
(\ref{eq:omega_1a}) through (\ref{eq:omega_3a}) 
[(\ref{eq:omega_1b}) through (\ref{eq:omega_3b})] do not tend towards 
steady rotation about $\ahat_2$.  Rather, they tend towards a state that 
oscillates between steady rotation with $\omega_2 = + J/I_2$ and 
$\omega_2 = - J/I_2$, with abrupt transitions between these two states of 
rotation.  This is illustrated in Figure \ref{fig:omega_flip_state}, 
where we plot $\omega_i$ for $q$ slightly less than $I_1/I_2$.  In Figure
\ref{fig:omega_flip_state}, ``$+$'' denotes the case with plus signs in 
equations (\ref{eq:omega_1a}) and (\ref{eq:omega_3a}) (the positive flip 
state with respect to $\ahat_1$; see \S \ref{sec:grain_flips}) and ``$-$''
denotes the case with minus signs in 
equations (\ref{eq:omega_1a}) and (\ref{eq:omega_3a}) (the negative flip 
state).  Note also that the period $P_{\tau} \rightarrow \infty$ as 
$q \rightarrow I_1/I_2$.  

\subsubsection{Grain Flips}
\label{sec:grain_flips}

From equations (\ref{eq:omega_1a}) and (\ref{eq:omega_1b}), we see that the 
grain state is not completely specified by $(\bJ, q)$.  
Given $\bJ$, there are two possible configurations for which a grain can 
spin about one of its principal axes $\ahat_i$---either $\ahat_i \parallel 
\bJ$ and $\omega_i >0$ or $\ahat_i \parallel - \bJ$ and $\omega_i < 0$.  We
will call the former configuration the ``positive flip state'' with 
respect to axis $\ahat_i$ and the latter configuration the ``negative flip 
state'' with respect to $\ahat_i$.  The flip states can be generalized
in a natural way to cases for which $\bJ$ does not lie along a principal axis.
When $q < I_1/I_2$, solutions of the Euler equations with a plus (minus)
sign in equation (\ref{eq:omega_1a}) have $\omega_1  > 0$ 
($\omega_1 < 0$),  
corresponding to a positive (negative) flip state with respect to axis 
$\ahat_1$.  (Also, $\omega_2$ and $\omega_3$ average to zero.)
When $q > I_1/I_2$, the plus (minus) sign in equation (\ref{eq:omega_1b})
corresponds to a positive (negative) flip state with respect to axis
$\ahat_3$.  

From equations (\ref{eq:omega_1a}), (\ref{eq:omega_3a}),
(\ref{eq:omega_1b}), and (\ref{eq:omega_3b}), we see that a grain can 
evolve smoothly from one flip state to another only when 
$\omega_1=0$ and $\omega_3=0$.  This only occurs when $q$ passes through
$I_1/I_2$.  Figure \ref{fig:omega_flip_state} shows $\omega_i$ for the 
two flip states about $\ahat_1$ when $q$ is slightly less than $I_1/I_2$. 
When $t/P_t = 0.25$, 0.75, 1.25, and 1.75, the $\omega_i$ are nearly identical 
for the two flip states and the motion is very similar to instantaneous 
rotation about $\ahat_2$.  Since $\omega_1$ does not quite reach zero when
$q < I_1/I_2$, these similarities are not quite exact.

\begin{figure}
\epsscale{1.00}
\plotone{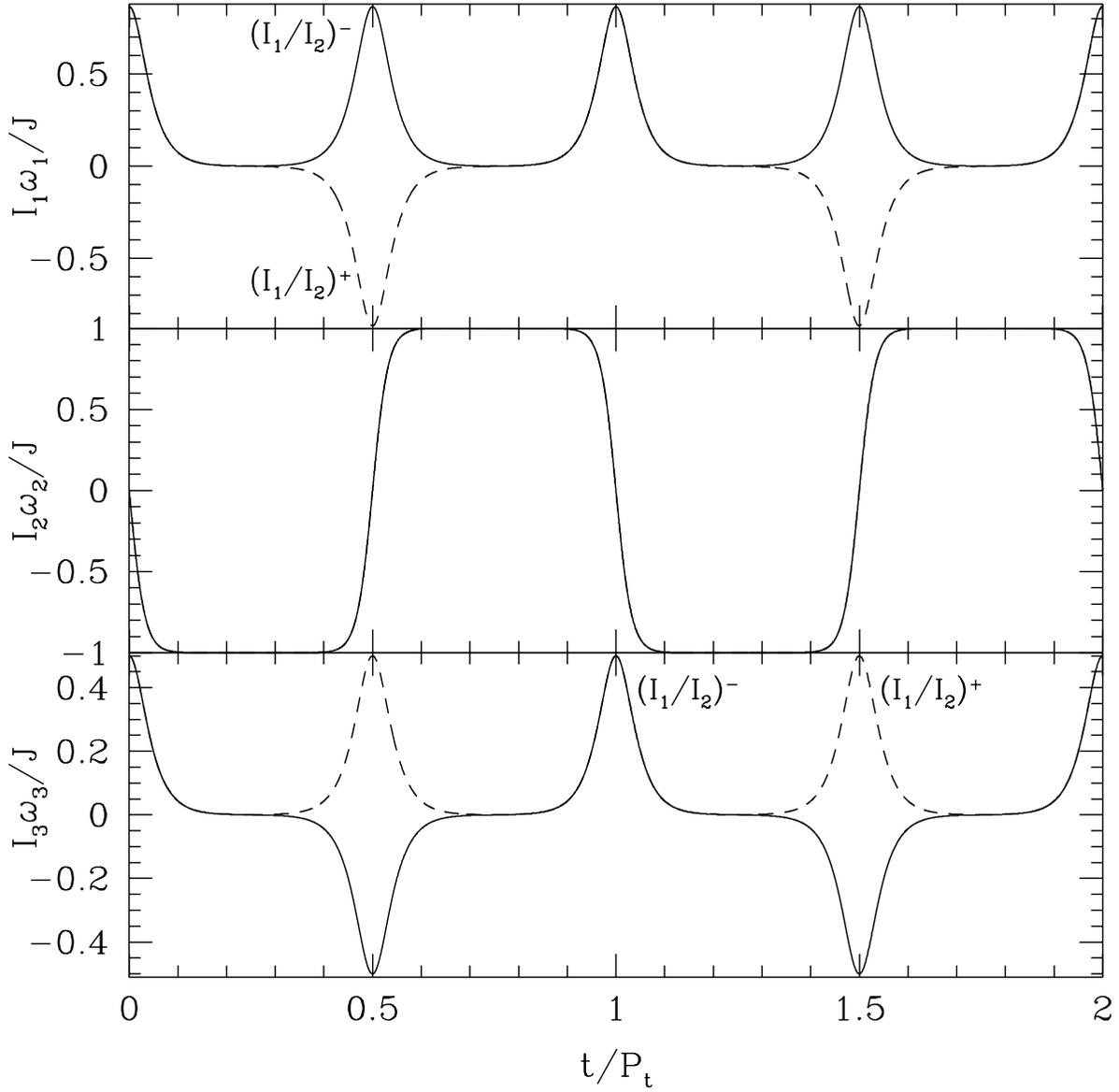}
\caption{
\label{fig:omega_flip}
Components of the angular velocity along grain principal axes, for 
$I_1:I_2:I_3 = 3:2:1$.  $(I_1/I_2)^+$ and $(I_1/I_2)^-$ indicate that $q$ is 
slightly larger or smaller than $I_1/I_2$, respectively.  The curves shown
are for the positive flip states with respect to $\ahat_1$ and $\ahat_3$.
The two curves are equivalent in regions where the dashed curve is absent.
        }
\end{figure}

In Figure \ref{fig:omega_flip} we plot
$I_i \omega_i/J$, both for $q$ slightly less than $I_1/I_2$ (solid curve)
and for $q$ slightly greater than $I_1/I_2$ (dashed curve).\footnote{Note
that the curve for $I_2 \omega_2/J$ is the same for both cases.}  
For these curves we have adopted $I_1:I_2:I_3=3:2:1$ and positive flip 
states.  Suppose the grain has
$q<I_1/I_2$ at $t=0$ and evolves along the solid curve
to $q=I_1/I_2$ when $t=P_t/4$.  At this
point, several different outcomes are possible.  If $q$ remains exactly 
equal to $I_1/I_2$, then $\omega_1$ and $\omega_3$ remain zero, and
$\omega_2 = -J/I_2$.  
If $q$ continues to increase, so that $q>I_1/I_2$, then there are two 
possibilities:  1. The 
$\omega_i$ emerge along the dashed curves rather than along the solid curves,
so that the grain is now in the positive flip state with respect to 
$\ahat_3$.  
2.  $\omega_1$ and $\omega_3$ emerge along the reflection of the 
dashed curves through the lines $\omega_i=0$, so that the grain is now in 
the negative flip state with respect to $\ahat_3$.  

Finally, suppose $q$ reaches a maximum value of $I_1/I_2$ at $t=P_t/4$ and
then decreases.  Again there are two possibilities:  1. The $\omega_i$ 
emerge along the solid curves, so that the grain remains in the positive 
flip state with respect to $\ahat_1$.  
2. $\omega_1$ and $\omega_3$ emerge along 
the reflection of the solid curves through the lines $\omega_i=0$.  This is
a grain flip; the grain is now in the negative flip state with respect to
$\ahat_1$
(and the angle $\gamma$ between $\bJ$ and $\ahat_1$ now exceeds $\pi/2$).
However, such a flip will occur with infinitesimal probability, since there
is infinitesimal probability that the maximum value of $q$ attained during
a fluctuation will be exactly $I_1/I_2$.  Real flips
will occur in a two-step process, in which the grain configuration evolves
across $q=I_1/I_2$ and then back again.

Recall that as $q \rightarrow I_1/I_2$, $P_t \rightarrow \infty$.  Thus, 
as $q$ gets very close to $I_1/I_2$, the evolution in $q$ occurs essentially
instantaneously compared with the timescale on which the $\omega_i$ change
for a given $q$.  The evolution to $q=I_1/I_2$
can occur at any time.  As $q$ changes, the curves
for the $\omega_i$ change, with $\omega_1 = 0$ and $\omega_3 = 0$ when 
$q=I_1/I_2$.  In the examples in this section, we have chosen to have the 
evolution to $q=0$ occur when $\omega_3 = 0$ and $\omega_1$ is nearly zero
for the initial $q$ (when $t=P_t/4$).  We made this choice so that only two 
curves would be needed in Figure \ref{fig:omega_flip}.  

In all cases, when $q$ crosses $q=I_1/I_2$, the grain is
equally likely to emerge in either flip state.

\subsubsection{\label{sec:axis_motion} Motion of Grain Axes}

Next, we need to find the motion of the grain axes in space.  The orientation 
of the grain axes in angular momentum coordinates can be expressed 
using Eulerian angles $(\alpha,\gamma,\zeta)$.
Suppose the grain axes $\ahat_2$, $\ahat_3$, $\ahat_1$ are 
originally lined up with $\xJ$, $\yJ$, $\zJ$.  Any grain
orientation can be obtained by performing the following operations on the
grain coordinates:  1.  Rotate through angle $\zeta$ about $\ahat_1 = \zJ$.
2.  Rotate through angle $\gamma$ about $\ahat_2$.  3.  Rotate through angle
$\alpha$ about $\ahat_1$.  (See Fig.~\ref{fig:coord_frames}.)

By resolving $\bJ$ onto grain axes, we find the
following relations for the Eulerian angles $\alpha$ and $\gamma$:
\be
\label{eq:axis_motion_1}
\sin \alpha \sin \gamma = I_2 \omega_2 / J~~~,
\ee
\be
\cos \alpha \sin \gamma = I_3 \omega_3 / J~~~,
\ee
and
\be
\label{eq:axis_motion_3}
\cos \gamma = I_1 \omega_1 / J~~~.
\ee
Angles $\alpha$ and $\gamma$ are periodic functions of time, with period $P_t$ 
(eqs.~\ref{eq:period1} and \ref{eq:period2}).  The remaining Eulerian angle,
$\zeta$, can be expressed as a sum of two periodic terms, one with 
period $P_t$, and the other with an incommensurate period.  Consequently,
the overall torque-free motion is not periodic.  However, given a function
$A(\alpha, \zeta, \gamma$), we can still find its average value 
$\bar{A}(q, \pm)$ over the torque-free motion, as follows:
\be
\label{eq:bar_A}
\bar{A}(q,\pm) = \frac{1}{P_{\tau}} \int_0^{P_{\tau}} d\tau \, \frac{1}{2 \pi} 
\int_0^{2 \pi} d\zeta \, A(\alpha, \zeta, \gamma)~~~.
\ee
In equation (\ref{eq:bar_A}), we have explicitly indicated the dependence 
of $\bar{A}(q, \pm)$ on $q$ and on the flip state; ``$+$'' (``$-$'') indicates
the positive (negative) flip state with respect to $\ahat_1$ or $\ahat_3$,
depending on $q$.  Of course, $\bar{A}$ may also depend on other variables
(e.g., $\bJ$), depending on which quantity $A$ is being averaged.

\subsubsection{Barnett Dissipation}
\label{sec:Barnett_diss}

A paramagnetic grain in steady rotation (constant $\bomega$) acquires a 
magnetization 
$\chi_0 \bomega/\gamma_{\rm g}$ (the Barnett effect), where $\chi_0$ is the 
static magnetic susceptibility and $\gamma_{\rm g}$ is the gyromagnetic ratio
of the microscopic magnetic dipoles that are responsible for the grain's
paramagnetism
($\gamma_{\rm g} = - g_e \mu_{\rm B}/\hbar \approx -1.76 \times 10^7 \s^{-1}
\, {\rm G}^{-1}$ for electrons and $\gamma_{\rm g}=g_{\rm N} \mu_{\rm N}/\hbar 
\approx 1.3 \times 10^4 \s^{-1} \, {\rm G}^{-1}$ for nuclei).
Thus, in describing the paramagnetic response of a grain, the vector 
${\bf B}_{\rm BE} \equiv \bomega / \gamma_{\rm g}$ may be regarded as an
effective magnetic field, termed the ``Barnett equivalent field'' by 
Purcell (1979).  

Purcell noted that when a paramagnetic grain undergoes 
non-steady rotation, ${\bf B}_{\rm BE}$ is not static in grain-body 
coordinates; consequently, rotational kinetic energy is dissipated.  
Purcell assumed that this dissipation rate is the same as the paramagnetic
dissipation rate for a static grain exposed 
to true magnetic fields given by $\bomega/\gamma_{\rm g}$, where 
$\bomega$ is evaluated in the (non-inertial!) reference frame in which the
grain is stationary.  
Purcell considered an axisymmetric grain and ignored
the constant component of $\bomega/\gamma_{\rm g}$ lying along the 
symmetry axis.  
Lazarian \& Draine (2000) showed that this component 
cannot be neglected without introducing serious error in the 
dissipation rate when the grain rotation is very fast. 
The Purcell prescription (sometimes modified as per Lazarian \& Draine)
for obtaining the rate at which rotational
energy is dissipated is widely used in the grain alignment literature, but
it has never actually been proved to be correct.  
Although errors are possible in some cases, we 
expect the prescription to be accurate when 
applied to the grains that polarize starlight.  

Since the angular momentum in the microscopic dipoles is generally
a tiny fraction of the total angular momentum of the grain, we make the 
idealization that Barnett dissipation does not affect the angular momentum
associated with the grain rotation; i.e., it affects the value of 
$q$ but not $\bJ$.  The role of the Barnett effect is to provide a coupling 
allowing energy exchange (at fixed angular momentum) between the rotational
and vibrational degrees of freedom of the lattice.

The Barnett equivalent field for a non-axisymmetric grain is quite 
complicated.  For example, suppose $q < I_1/I_2$.  Equations 
(\ref{eq:omega_1a}) through (\ref{eq:omega_3a}) reveal that, in this case:
(a) The biasing field along $\ahat_1$ is not constant; it has a 
non-sinusoidally oscillating component. 
(b) The rotating component does not consist of a single field with 
constant magnitude executing sinusoidal rotation.  
A correct description of the paramagnetic dissipation rate for this case
would be prohibitively complicated.

However, a detailed analysis of the Barnett dissipation rate would not
be warranted, because:  (a) The paramagnetic properties of interstellar
grains are poorly known, so we cannot estimate the dissipation rate with 
high accuracy.  (b) As we will see,
the Barnett dissipation timescale is many orders of magnitude shorter 
than the timescales for some of the other processes that play important
roles in the grain dynamics.  Thus, we will not strictly follow the 
evolution of $q$ due to Barnett dissipation and fluctuations, but will 
instead adopt a simple approximate approach, as described below 
(\S \ref{sec:strategy_just_flips}).  In this approach, we will simply 
require an estimate of the Barnett dissipation timescale $\tau_{\rm Bar}$.

In estimating $\tau_{\rm Bar}$, we will consider the case of
an axisymmetric (specifically, oblate; i.e., $I_1 > I_2 = I_3$) grain.
We will derive an order of magnitude estimate of $\tau_{\rm Bar}$ and 
assume that this estimate also holds for non-axisymmetric grains.
For oblate grains, 
${\bf B}_{\rm BE}$ consists of a static component $(\omega_{\parallel}/
\gamma_g) \ahat_1$ plus a component $\omega_{\perp}/\gamma_g$
that rotates in the $\ahat_1 - \ahat_2$ plane, with angular velocity
$\omega_{\rm rot} \, \ahat_1$.  Setting $I_2=I_3$ in equations 
(\ref{eq:omega_1a}) through (\ref{eq:period1}), we find
\be
\label{eq:omega_parallel_oblate}
\omega_{\parallel} = \frac{J}{I_1} \cos \gamma~~~,
\ee
\be
\omega_{\perp} = \frac{J}{I_3} \sin \gamma~~~,
\ee
and 
\be
\omega_{\rm rot} = \frac{J(I_1-I_3)}{I_1 I_3} \cos \gamma~~~,
\ee
where $\gamma$ is the (constant) angle between $\bJ$ and $\ahat_1$.
The energy dissipation rate is given by
\be
\label{eq:Barnett_diss}
\left( \frac{dE}{dt} \right)_{\rm Bar} = - V \chi^{\prime \prime}
B_{\rm BE, \, rot}^2 \omega_{\rm rot}~~~,
\ee
where $V$ is the grain volume, $\chi^{\prime \prime}$ is the
imaginary component of the magnetic susceptibility, and
$B_{\rm BE, \, rot} \equiv \omega_{\perp}/\gamma_g$ is the rotating 
component of the Barnett equivalent field.

The susceptibility can be roughly approximated using the solution of the 
modified Bloch equations derived by Wangsness (1956, his eq.~A13):
\be
\label{eq:chi_imag}
\chi^{\prime \prime} = \frac{\chi_0 \omega_{\rm rot} T_2}{1+
(\omega_{\parallel} + \omega_{\rm rot})^2 T_2^2 + 
\omega_{\perp}^2 T_1 T_2}~~~,
\ee
where $T_1$ is the spin-lattice relaxation time and $T_2$ is the 
spin-spin relaxation time.  For electron paramagnetism, we take 
$\chi_0 T_2 \sim 10^{-13} (15\K/T_d) \s$, $T_2 \sim 3\times 
10^{-11}$--$3\times 10^{-9} \s$, and $T_1 \sim 10^{-6} \s$ (Draine 1996).
For nuclear paramagnetism, we take $\chi_0 \sim 4 \times 10^{-11}
(15 \K /T)$, $T_2 \sim 3\times 10^{-5}$--$3\times 10^{-4} \s$, and 
$T_1 \sim T_2$ (Lazarian \& Draine 1999b).

For an oblate grain,
\be
\label{eq:q_oblate}
q = 1+ \frac{I_1-I_3}{I_3} \sin^2 \gamma~~~.
\ee
From equations (\ref{eq:omega_parallel_oblate}) through (\ref{eq:q_oblate}), 
we find that (for $J$ held constant, as we assume to be the case for 
Barnett dissipation)
\be
\frac{dq}{dt} = - \tau_{\rm Bar}^{-1} (q-1) \frac{(1 - q I_3/I_1)}{(1-I_3/I_1)}
~~~,
\ee
\be
\tau_{\rm Bar}^{-1}\equiv\frac{2 V \chi_0 T_2}{\gamma_{\rm g}^2D} 
\frac{J^2(I_1-I_3)}{I_1I_3^3}
= \frac{2 V \chi_0 T_2}{\gamma_{\rm g}^2I_3D} 
\frac{(I_1-I_3)}{I_3} ~ \omega^2~~~,
\ee
where $\omega^2\equiv J^2/(I_1I_3)$ and
\be
\label{eq:D}
D = 1 + \frac{J^2}{I_3^2} (\cos^2 \gamma T_2^2 + \sin^2 \gamma T_1 T_2)
\sim 1 + \frac{I_1}{2I_3}\omega^2 T_1 T_2~~~,
\ee
where we have taken
$\sin^2 \gamma \approx \cos^2 \gamma \approx 1/2$ .
Thus,
\be
\tau_{\rm Bar}^{-1} \sim \frac{2 V \chi_0 T_2}{\gamma_{\rm g}^2 I_3}
\frac{(I_1-I_3)}{I_3}
\frac{\omega^2}{1 + (I_1/2I_3) \omega^2 T_1 T_2}~~~.
\ee
The Barnett energy dissipation rate 
$(dE/dt)_{\rm Bar} \sim - (J^2 / 2 I_1) \tau_{\rm Bar}^{-1}(q-1)
(1-q I_3/I_1)/(1-I_3/I_1)$.  Note that $\tau_{\rm Bar}$ is the exponential
decay time for $q-1$ when $q-1 \ll 1$.  

Adding the dissipation rates resulting from both electron and nuclear
paramagnetism, we estimate
\begin{eqnarray}
\label{eq:tau_Bar}
\tau_{\rm Bar}^{-1} & \sim & 1.7 \yr^{-1} \omega_5^2
\left(\frac{0.1\micron}{a}\right)^2 \left(\frac{15\K}{T_d}\right)
\left( \frac{3 \g \cm^{-3}}{\rho} \right) \left(\frac{I_1-I_3}{I_3}\right)
\left[ \frac{1}{1+3\times 10^{-6} \omega_5^2} + 
\frac{7\times 10^4}{1+ 100 \omega_5^2}\right] \nonumber
\\
& \sim & 
4.7 \yr^{-1} 
\left(\frac{\omega}{\omega_{\rm T}}\right)^2 
\left(\frac{0.1\micron}{a}\right)^7 
\left( \frac{3 \g \cm^{-3}}{\rho} \right)^2 
\left(\frac{15\K}{T_d}\right)
\left( \frac{\Tgas}{100 \K} \right)
\left(\frac{I_1-I_3}{I_3}\right) \times
\nonumber
\\
&&\left[ \frac{1}{1+8\times 10^{-6} p} + 
\frac{7\times 10^4}{1+ 280 p}\right]~~~,
\end{eqnarray}
where $\omega_5 = \omega / 10^5 \s^{-1}$ and 
\be
p = \left(\frac{\omega}{\omega_{\rm T}}\right)^2 
\left( \frac{3 \g \cm^{-3}}{\rho} \right)
\left( \frac{T_{\rm gas}}{100 \K} \right)
\left(\frac{0.1\micron}{a}\right)^5~~~.
\ee
Note the very steep dependence of the dissipation rate on $a$ when the 
grain rotates thermally (recall eq.~\ref{eq:omega_T} for the thermal 
rotation speed $\omega_{\rm T}$).
For large $a$ (for which $p$ is small), we expect dissipation 
due to nuclear paramagnetism to dominate that due to electron 
paramagnetism, as pointed out by Lazarian \& Draine (1999b).
In Figure \ref{fig:timescales}, we plot $\tau_{\rm Bar}$ versus $a$ 
for $\omega = \omega_{\rm T}$ and assuming $I_1 \approx 2 I_3$.

\section{\label{sec:torques} Torques}

\subsection{Coordinate Transformation}

In order to find values of $\bar{\bGam}(q,\pm)$ and 
$\overline{\bGam \cdot \bomega}(q,\pm)$
for the various torques acting on a grain, we need to transform vectors from 
grain body coordinates to angular momentum coordinates.  The transformation
is a sequence of three rotations, and is given by the following equations:
\be
\xJ = (\cos\alpha \cos\zeta - \sin\alpha \sin\zeta \cos\gamma) \,
\ahat_2 - (\sin\alpha \cos\zeta + \cos\alpha \sin\zeta \cos\gamma) \,
\ahat_3 + \sin\zeta \sin\gamma \, \ahat_1~~~,
\ee
\be
\yJ = (\cos\alpha \sin\zeta + \sin\alpha \cos\zeta \cos\gamma) \,
\ahat_2 + (\cos\alpha \cos\zeta \cos\gamma - \sin\alpha \sin\zeta) \,
\ahat_3 - \cos\zeta \sin\gamma \, \ahat_1~~~,
\ee
and
\be
\zJ = \sin\alpha \sin\gamma \, \ahat_2 + \cos\alpha \sin\gamma \,
\ahat_3 + \cos\gamma \, \ahat_1~~~.
\ee

\subsection{Barnett Torque}
 
The Barnett torque ${\bGam}_{\rm B} = \bmu_{\rm Bar} \times \bB$.  
For a grain in steady rotation, the Barnett magnetic moment is given by
\be
\label{eq:mu_Bar_uniform_rotation}
\bmu_{\rm Bar} = \frac{\chi_0 V}{\gamma_{\rm g}} \bomega~~~;
\ee
for non-steady rotation, the magnetization is more complicated, as discussed
in \S \ref{sec:Barnett_diss}.  If we assume that equation 
(\ref{eq:mu_Bar_uniform_rotation}) holds even for the case of non-steady 
rotation, then $\bar{\bGam}_{\rm B}$ can be found very easily. 
Since $\overline{\bomega \times \bB} = \bar{\bomega} \times \bB$, 
$\bar{\bGam}_{\rm B} = (\chi_0 V/\gamma_{\rm g}) \bar{\bomega} \times \bB$.
Transforming $\bomega$ from grain body coordinates to angular momentum 
coordinates and averaging over $\zeta$, we find that
\be
\label{eq:omega_ast}
\bar{\bomega}(q) = q \bJ / I_1~~~;
\ee
note that $\bar{\bomega}$ does not depend on the flip state.
Thus, 
\be
\label{eq:Gamma_Bar_ast}
\bar{\bGam}_{\rm B}(q) = \hatphi \, q J \Omega_{\rm B} \sin\xi~~~,
\ee
where
\be
\label{eq:Omega_B}
\Omega_{\rm B} = - \frac{\chi_0 V B}{\gamma_{\rm g} I_1} \approx
25 \yr^{-1} \left( \frac{0.1 \micron}{a} \right)^2
\left( \frac{3 \g \cm^{-3}}{\alpha_1 \rho}\right) \left( \frac{\chi_0}
{3.3 \times 10^{-4}} \right) \left( \frac{B}{5 \mu G} \right)~~~.
\ee
Since $\Omega_{\rm B} \propto \chi_0$, electron paramagnetism dominates 
nuclear paramagnetism here.  A correct treatment of the magnetization for 
the case of non-steady grain rotation does not significantly modify the 
above result.  For example, for an oblate axisymmetric grain, equation 
(\ref{eq:Gamma_Bar_ast}) need only be modified by the 
substitution\footnote{We used the susceptibilities given by Wangsness
(1956, his eqs.~A11 through A13) in deriving the substitution 
(\ref{eq:q_substitution}).}
\be
\label{eq:q_substitution}
q \rightarrow q + (q-1) x \, \frac{1+ x \left( 2 \frac{I_2}{I_1} + \frac{T_1}
{T_2} -1 \right)}{1 + x^2 \left[ \left( 2 \frac{I_2}{I_1} -1 \right)^2
+ \frac{I_2 (q-1)}{I_1 - I_2 q} \frac{T_1}{T_2} \right]}~~~,
\ee
where
\be
x \equiv \left( \frac{I_1 - I_2 q}{I_1 - I_2} \right)^{1/2} \frac{J}{I_2}
T_2~~~.
\ee
The correction term can be safely ignored since $x$ only becomes significant
when the grain rotation is highly suprathermal, in which case $q-1$ is small.
Note that the timescale for precession about $\bB$ is rapid compared with the 
timescales for all other relevant processes, except for the grain rotation
itself and possibly the Barnett dissipation timescale.  

Since $\bGam_{\rm B} \cdot \bomega = (\bmu_{\rm Bar} \times \bB) \cdot
\bomega = (\bomega \times \bmu_{\rm Bar}) \cdot \bB$, we need to consider
the component of $\bmu_{\rm Bar}$ that lags $\bomega$ in order
to find $\overline{\bGam_{\rm B} \cdot \bomega}$.  The relevant component 
is $\mu_{\rm Bar} \sim \chi^{\prime \prime} \omega V/\gamma_g$; thus,
$\overline{\bGam_{\rm B} \cdot \bomega} \sim \chi^{\prime \prime} \omega^2
B V/\gamma_g$.  This is identical to $(dE/dt)_{\rm Bar}$ except that the
interstellar field $B$ is substituted for the Barnett equivalent field
$\omega/\gamma_g$.  Thus, 
\be
\label{eq:Gamma_B_dot_omega}
\overline{\bGam_{\rm B} \cdot \bomega} \sim \frac{\gamma_{g, \, e^-} B}
{\omega} \left( \frac{dE}{dt} \right)_{{\rm Bar}, \, e^-} + 
\frac{\gamma_{g, \, {\rm nuc}} B}
{\omega} \left( \frac{dE}{dt} \right)_{\rm Bar, \, nuc}~~~,
\ee
where the subscripts ``$e^-$'' and ``nuc'' denote electron and nuclear
paramagnetism, respectively.  At the low rotational speeds that may be
encountered for large grains during crossovers, the electron term in 
equation (\ref{eq:Gamma_B_dot_omega}) may be comparable to the electron 
Barnett dissipation rate, but in this case nuclear Barnett dissipation 
dominates $\overline{\bGam_{\rm B} \cdot \bomega}$.  

\subsection{\label{sec:drag} Drag}

Collisions with gas atoms will exert a drag torque, which can be
written
\be
\bar{\bGam}_{\rm drag, \, gas} \approx - \bJ / \tau_{\rm drag, \, gas}~~~,
\ee
where, for a sphere,
\be
\label{eq:drag_time}
\tau_{\rm drag, \, gas} = 8.72 \times 10^4 \yr \left(\frac{\rho}{3 \g
\cm^{-3}} \right) \left( \frac{a}{0.1 \micron} \right) \left( \frac{\Tgas}
{100 \K} \right)^{1/2} \left( \frac{3000 \cm^{-3} \K}{\nH \Tgas} \right)~~~.
\ee
For nonspherical shapes, the drag time will tend to be somewhat larger
than given by eq.\ (\ref{eq:drag_time}), but only
by factors of order $\sim1.5$.
The drag torque will also depend on the orientation of the
grain relative to $\bJ$, but this dependence is quite weak and will
be neglected.

There is also a drag torque due to the emission of infrared photons by
the grain.  This torque has the same form as the gas drag torque but is
characterized by a drag timescale 
\be
\label{eq:tau_drag_em}
\tau_{\rm drag, \, em} = 1.1 \times 10^5 \yr \,
\langle \Qabs \rangle^{-1} \left( \frac{\rho}{3 \g \cm^{-3}} \right)
\left( \frac{a}{0.1 \micron} \right)^3 \left( \frac{T_d}{15 \K} \right)^2
\left( \frac{u_{\rm ISRF}}{u_{\rm rad}} \right)
\ee
(Paper I).  In equation (\ref{eq:tau_drag_em}), $u_{\rm rad}$ is the 
energy density of the incident radiation field (that heats the grain)
and $u_{\rm ISRF} = 8.64 \times 10^{-13} \erg \cm^{-3}$ is the energy 
density in the average interstellar radiation field (ISRF) in the solar
neighborhood, as determined by Mezger, Mathis, \& Panagia (1982) and
Mathis, Mezger, \& Panagia (1983).  The grain absorption cross section
is $\pi a^2 \Qabs(a, \lambda)$ and the angle brackets denote averaging 
over the incident spectrum.  

The total drag timescale is given by
\be
\tau_{\rm drag}^{-1} = \tau_{\rm drag, \, gas}^{-1} + 
\tau_{\rm drag, \, em}^{-1}~~~.
\ee

\begin{figure}
\epsscale{1.00}
\plotone{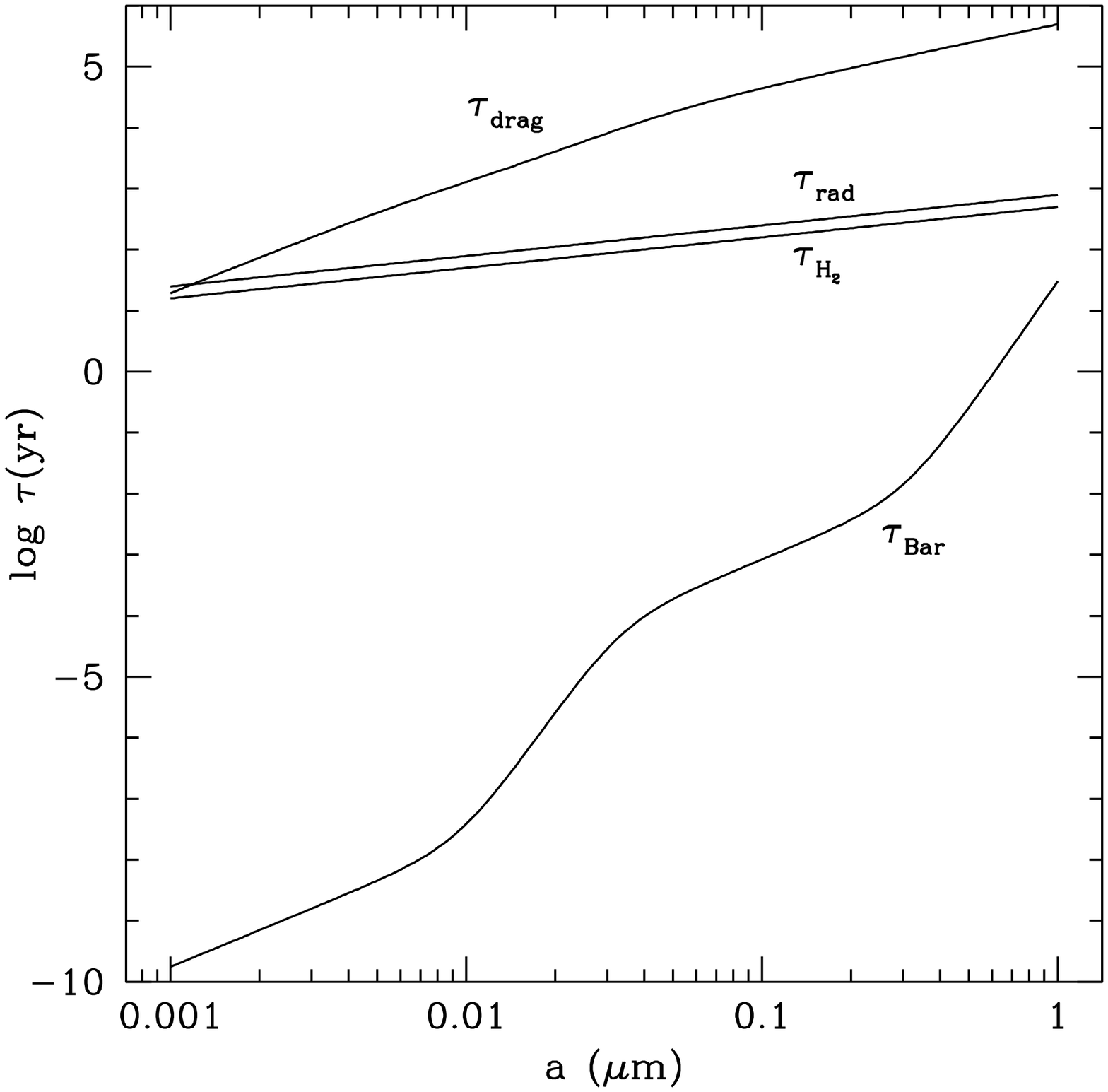}
\caption{
\label{fig:timescales}
Timescales for Barnett relaxation ($\tau_{\rm Bar}$), the drag torque
($\tau_{\rm drag}$), the H$_2$ formation torque ($\tau_{{\rm H}_2}$), and
radiative torques ($\tau_{\rm rad}$) for silicate grains,
assuming $\omega = \omega_{\rm T}$, $I_1 \approx 2 I_3$, 
$T_d = 15 \K$, $\rho = 3 \g \cm^{-3}$, $T = 100\K$, $\nH = 30 \cm^{-3}$,
$u_{\rm rad} = u_{\rm ISRF}$, $f=1$, $l=10 \Angstrom$, 
$E_{{\rm H}_2} = 0.2 \eV$, $n({\rm H}) = \nH$, and $\gamma_{\rm rad} H
= 10^{-3}$.  Both $\tau_{{\rm H}_2}$ and $\tau_{\rm rad} \propto a^{0.5}$
and, coincidentally, are nearly identical for the above canonical conditions.
        }
\end{figure}

Also, 
\be
\overline{\bGam_{\rm drag} \cdot \bomega} = - I_1 
\overline{\omega^2} / \tau_{\rm drag}~~~,
\ee
which implies
\be
\frac{\overline{\bGam_{\rm drag} \cdot \bomega}}
{(dE/dt)_{\rm Bar}} \sim \frac{\tau_{\rm Bar}}{\tau_{\rm drag}}~~~.
\ee
Figure \ref{fig:timescales} shows $\tau_{\rm drag}$ versus $a$.  In 
computing $\langle \Qabs \rangle$, we have assumed spherical grains, 
using Mie theory and the dielectric functions for silicates from Li \&
Draine (2001).  The specific energy density $u_{\lambda}$ for the ISRF
is given in Paper I.
From Figure \ref{fig:timescales} we see that 
$\tau_{\rm Bar}/\tau_{\rm drag} \ll 1$ for all grain sizes of 
interest when $\omega \sim \omega_{\rm T}$.  Thus, 
$\overline{\bGam_{\rm drag} \cdot \bomega}$
may be neglected in the evolution of $q$. 

\subsection{\label{sec:H2} H$_2$ Formation}

The H$_2$ formation torque $\bGam_{{\rm H}_2}$ has a fixed direction in 
grain body coordinates.  When $q< I_1/I_2$,
\be
\bar{\bGam}_{{\rm H}_2}(q,\pm) = 
\pm \left( \bGam_{{\rm H}_2} \cdot \ahat_1 \right)
\left[ \frac{I_1 -I_3 q}{I_1 - I_3} \right]^{1/2} \frac{\pi}{2 F(\pi/2, k^2)}
\hatJ
\ee
and
\be 
\label{eq:dot_H_2_1}
\overline{\bGam_{{\rm H}_2} \cdot \bomega}(q,\pm) = \pm 
| \bar{\bGam}_{{\rm H}_2} | \, J/I_1~~~.
\ee
When $q> I_1/I_2$,
\be
\bar{\bGam}_{{\rm H}_2} (q,\pm) = 
\pm \left( \bGam_{{\rm H}_2} \cdot \ahat_3 \right)
\left[ \frac{I_3 (q-1)}{I_1 -I_3} \right]^{1/2} \frac{\pi}{2 F(\pi/2, k^{-2})}
\hatJ
\ee
and
\be
\label{eq:dot_H_2_2}
\overline{\bGam_{{\rm H}_2} \cdot \bomega}(q,\pm) = 
\pm | \bar{\bGam}_{{\rm H}_2} | \, J/I_3~~~.
\ee

We estimate the magnitude of the H$_2$ formation torque as (see Paper II)
\be
\label{eq:H_2_mag}
\Gamma_{{\rm H}_2} = I_1 \omega_{{\rm H}_2}/\tau_{\rm drag, \, gas}~~~,
\ee
with
\be
\omega_{{\rm H}_2} \sim 5.2 \times 10^7 \s^{-1} f \left(
\frac{0.1 \micron}{a} \right)^2 \left( \frac{l}{10 \Angstrom} \right) \left(
\frac{E_{{\rm H}_2}}{0.2 \eV} \right)^{1/2} \left[ \frac{n({\rm H})}
{\nH} \right]~~~;
\ee
$f$ is the fraction of the H atoms that collide with the grain and depart
as H$_2$, $l^2$ is the surface area per H$_2$ formation site on the grain 
surface, $E_{{\rm H}_2}$ is the kinetic energy of the departing H$_2$ 
molecules, and $n({\rm H})/\nH$ is the fraction of the gas-phase H that is in
atomic form.  For a real grain, $\bGam_{{\rm H}_2}$ varies in magnitude and
direction on a timescale characteristic of changes in the surface sites.
As in Paper II, we will ignore this complication and assume that 
$\bGam_{{\rm H}_2}$ does not evolve with time.

Equations (\ref{eq:dot_H_2_1}), (\ref{eq:dot_H_2_2}), and 
(\ref{eq:H_2_mag}) yield
\be
\frac{\overline{\bGam_{{\rm H}_2} \cdot \bomega}}{(dE/dt)_{\rm Bar}} \sim 
\frac{\tau_{\rm Bar}}{\tau_{\rm drag, \, gas}} \frac{\omega_{{\rm H}_2}}
{\omega}~~~.
\ee
Figure \ref{fig:timescales} shows that $\tau_{{\rm H}_2} \equiv 
\tau_{\rm drag, \, gas} \omega / \omega_{\rm T} \gg \tau_{\rm Bar}$ 
when $\omega \sim \omega_{\rm T}$, so 
$\overline{\bGam_{{\rm H}_2} \cdot \bomega}$ may be neglected in the 
evolution of $q$. 

\subsection{Paramagnetic Dissipation}

For steady grain rotation, the torque due to paramagnetic dissipation is 
given by (Davis \& Greenstein 1951)
\be
\bGam_{\rm DG} = \frac{\chi^{\prime \prime} V}{\omega} \left(
\bomega 
{\bf \times} \bB \right) {\bf \times} \bB~~~.
\ee
Lazarian \& Draine (2000) pointed out that $\chi^{\prime \prime}$
is modified by the grain rotation itself.  They showed that the dissipation
rate for a sample rotating in a weak static field $\bB$ is equal to the 
dissipation rate for a static sample subject to the Barnett equivalent field 
$\bomega / \gamma_g$ plus the rotating field $\bB$.  This yields 
\be
\label{eq:chi_imag_DG}
\chi^{\prime \prime} = \frac{\chi_0 \omega T_2}{1 + \gamma_{\rm g}^2
B^2 \sin^2 \xi T_1 T_2} \sim \chi_0 \omega T_2~~~;
\ee
the second term in the denominator is negligible for the magnetic field
strengths encountered in the ISM.  Since $\chi_0 T_2$ for electron 
paramagnetism substantially exceeds that for nuclear paramagnetism, the
nuclear contribution may be neglected.
Note that equation (\ref{eq:chi_imag_DG})
does not have the same form as equation (\ref{eq:chi_imag}) for 
$\chi^{\prime \prime}$ in the case of Barnett dissipation.  The 
rotation of the interstellar magnetic field (in the grain's frame) is
in the correct sense so as to be in resonance with the Barnett 
equivalent field lying along $\ahat_1$, so the term involving 
$T_2^2$ is zero.  The rotating component of the Barnett equivalent 
field, on the other hand, rotates in the wrong direction for resonance.

The situation is much more complicated in the case of non-steady grain 
rotation, due to the increased complexity of the external field as viewed
in the grain's frame.  For example, for an axisymmetric grain in 
non-steady rotation, the external field consists of five separate 
components: a constant component along $\ahat_1$, a component that 
oscillates along $\ahat_1$, and three components that rotate (with 
different angular speeds) about $\ahat_1$.  Even the simplest analysis,
in which the components of the external field along $\ahat_1$ are 
neglected in comparison with the Barnett equivalent field 
$\omega_1 \ahat_1 / \gamma_g$, introduces a complicated dependence on 
$q$, since the field strengths and angular frequencies of the three 
rotating components all depend on $q$.\footnote{It should also be noted
        that the susceptibilities are not necessarily the same as for 
        the case of a single rotating field component.  The modified 
        Bloch equations should be solved with all of the rotating field
        components included.  It seems unlikely that the susceptibilities
        would be significantly affected when all of the rotating fields
        are as weak as the interstellar field.  However, there could be
        cases in which the presence of the rotating components of the 
        Barnett equivalent field affects the susceptibilities for the 
        components of the interstellar field.}  

Since we are primarily concerned with radiative torques in this paper,
we will ignore the complicated $q$-dependence of the Davis-Greenstein 
torque, and simply treat it as if the grain were rotating uniformly 
about $\ahat_1$.  In this case, 
\be
\label{eq:D-G_torque}
\bar{\bGam}_{\rm DG} = 
- \frac{J}{\tau_{\rm DG}} \sin\xi \left(
\hatxi \cos\xi + \hatJ \sin\xi \right)~~~,
\ee
with
\be
\tau_{\rm DG} = \frac{2 \alpha_1 \rho a^2}{5 \chi_0 T_2 B^2} 
\sim 1.5 \times 10^6 \yr
\left( \frac{\alpha_1 \rho}{3 \g \cm^{-3}} \right) \left( \frac{a}{0.1
\micron} \right)^2 \left( \frac{T_d}{15 \K} \right) \left( \frac{5
\mu {\rm G}}{B} \right)^2~~~.
\ee

Since $\overline{\bGam_{\rm DG} \cdot \bomega} \sim \chi^{\prime \prime}
V \omega B^2$, $\overline{\bGam_{\rm DG} \cdot \bomega}$ is suppressed
with respect to $(dE/dt)_{\rm Bar}$ by a factor $\sim (B \gamma_{\rm g}/
\omega)^2 (1 + \omega^2 T_1 T_2)$, rendering it negligible. 

\subsection{Radiative Torques}

The radiative torque on a stationary grain exposed to a unidirectional 
radiation field with energy density $u$ can be expressed as
(Paper I)
\be
\label{eq:Gamma_rad}
\bGam_{\rm rad} = \pi a^2 u \frac{\tilde{\lambda}}{2 \pi}
\langle \bQ_{\Gamma} (\Theta, \Phi, \beta) \rangle~~~,
\ee
where a tilde denotes spectral averaging:
\be
u \equiv \int u_{\lambda}~d\lambda~~~,
\ee
\be
\label{eq:tilde_lambda}
\tilde{\lambda} \equiv \frac{1}{u} \int \lambda~u_\lambda~d\lambda
\ee
and
\be
\label{eq:tilde_Q}
\tilde{\bQ}_{\Gamma} (\Theta, \Phi, \beta) \equiv \frac{1}
{\tilde{\lambda} u} \int 
\bQ_{\rm \Gamma} (\Theta, \Phi, \beta, \lambda)~
\lambda~ u_\lambda ~d\lambda ~~~.
\ee

We approximate the interstellar radiation field by an isotropic component
with energy density $(1-\gamma_{\rm rad}) u_{\rm rad}$ plus a 
unidirectional component with energy density $\gamma_{\rm rad} u_{\rm rad}$,
propagating at angle $\psi$ with respect to $\bB$.  
We will assume that the starlight anisotropy direction is in the
$\xB-\zB$ plane.

\subsubsection{Unidirectional Component}
\label{sec:radtorque_uni}

We adopt the following transformation between ``alignment coordinates''
and ``scattering 
coordinates'' for the unidirectional component of the radiation field:
\be
\label{eq:ehat_1}
\ehat_1 = - \sin\psi \, \xB + \cos\psi \, \zB~~~,
\ee
\be
\ehat_2 = \cos\psi \, \xB + \sin\psi \, \zB~~~,
\ee
\be
\label{eq:ehat_3}
\ehat_3 = \yB~~~;
\ee
see Figure \ref{fig:coord_frames}.

The unidirectional component of the radiative torque is then given by
\be
\bar{\bGam}_{\rm rad}^{\rm uni}(\xi, \phi, \psi, q, \pm) = \frac{1}{2} 
\gamma_{\rm rad} u_{\rm rad} a^2 \tilde{\lambda} 
\left[ F(\xi, \phi, \psi, q, \pm) \hatxi + 
G(\xi, \phi, \psi, q, \pm) \hatphi + H(\xi, \phi, \psi, q, \pm) \hatJ 
\right]~~~,
\ee
where  
\begin{eqnarray}
\label{eq:F}
F(\xi, \phi, \psi, q, \pm) &=&
 - \Qspec \cdot \ehat_1 \left( \sin \psi \cos \xi \cos \phi 
+ \cos \psi \sin \xi \right) \nonumber \\ &&
+  \Qspec \cdot \ehat_2 \left(
\cos \psi \cos \xi \cos \phi - \sin \psi \sin \xi \right) 
+ \Qspec \cdot \ehat_3 \cos \xi \sin \phi~~~, 
\end{eqnarray}
\be
\label{eq:G}
G(\xi, \phi, \psi, q, \pm) = \Qspec \cdot \ehat_1 \sin \psi \sin \phi 
- \Qspec \cdot \ehat_2 \cos \psi \sin \phi + \Qspec \cdot \ehat_3 
\cos \phi~~~,
\ee
and
\begin{eqnarray}
\label{eq:H}
H(\xi, \phi, \psi, q, \pm) &=&
\Qspec \cdot \ehat_1 \left( \cos \psi \cos \xi - \sin \psi \sin \xi \cos \phi
\right) \nonumber \\ &&
+ \Qspec \cdot \ehat_2 \left( \sin \psi \cos \xi + \cos \psi \sin \xi \cos
\phi \right) + \Qspec \cdot \ehat_3 \sin \xi \sin \phi~~~.
\end{eqnarray}
Note that equations (\ref{eq:F}) through (\ref{eq:H}) are identical to 
equations (21) through (23) in Paper II, except that here the efficiency
factor $\Qspec$ is averaged over the torque-free motion described in 
\S \ref{sec:torque-free}, whereas in Paper II the efficiency factors are
averaged over rotation about $\ahat_1$.  
In order to evaluate $\Qspec(\xi, \phi, \psi, q, \pm)$, we need to know
the values of $(\Theta, \Phi, \beta)$ that correspond to the Eulerian 
angles $(\alpha, \zeta, \gamma)$ as the grain executes its torque-free motion.
For this, we use equations (\ref{eq:Theta}) through (\ref{eq:beta}); 
expressions for the dot products appearing in these equations are given in
Appendix B.

For unidirectional radiation,
\be
\overline{\bGam_{\rm rad} \cdot \bomega} (\xi, \phi, \psi, q, \pm) =
\frac{1}{2} \gamma_{\rm rad} u_{\rm rad} a^2 \tilde{\lambda}
\frac{q J}{I_1} H(\xi, \phi, \psi, q, \pm)~~~.
\ee
Thus, 
\be
\frac{\overline{\bGam_{\rm rad} \cdot \bomega}}{(dE/dt)_{\rm Bar}}
\sim \frac{\tau_{\rm Bar}}{\tau_{\rm rad}}~~~,
\ee
where
\be
\tau_{\rm rad} = 250 \yr \left( \frac{\rho}{3 \g \cm^{-3}} \right)^{1/2}
\left( \frac{T}{100 \K} \right)^{1/2} \left( \frac{a}{0.1 \micron} 
\right)^{1/2} \left( \frac{10^{-3}}{\gamma_{\rm rad} H} \right)
\left( \frac{u_{\rm ISRF}}{u_{\rm rad}} \right) 
\left( \frac{1.2 \micron}{\tilde{\lambda}} \right) 
\left( \frac{\omega}{\omega_{\rm T}} \right)~~~.
\ee
We expect $\gamma_{\rm rad} H \sim 10^{-3}$ (Paper II).  For the ISRF, 
$\tilde{\lambda} = 1.2 \micron$.  Figure \ref{fig:timescales} shows that 
$\tau_{\rm Bar}/\tau_{\rm rad} \ll 1$ when $\omega \sim \omega_{\rm T}$, 
so $\overline{\bGam_{\rm rad} \cdot \bomega}$ may be neglected in evolving 
$q$.  

\subsubsection{Isotropic Component}
\label{sec:radtorque_iso}

The torque on a grain exposed to the isotropic component of the radiation 
field is fixed in grain body coordinates and is given by
\be
\bGam_{{\rm rad}, 0}^{\rm iso} 
= \frac{1}{2} \left( 1 - \gamma_{\rm rad} \right) u_{\rm rad} a^2
\tilde{\lambda} \tilde{\bQ}_{\Gamma}^{\rm iso}~~~,
\ee
where
\be
\tilde{\bQ}_{\Gamma}^{\rm iso} \cdot \ahat_i = \frac{1}{4\pi}
\int_{-1}^1 d\cos\Theta \int_0^{2\pi} d\beta \tilde{\bQ}_{\Gamma}(\Theta,
0, \beta) \cdot \ahat_i~~~.
\ee

As with the H$_2$ formation torque, the components perpendicular to 
$\bJ$ average to zero during the grain rotation.  Thus,
\be
\bar{\bGam}_{\rm rad}^{\rm iso}(q,\pm) = 
\pm \left( \bGam_{{\rm rad}, 0}^{\rm iso} \cdot \ahat_1 \right)
\left[ \frac{I_1 -I_3 q}{I_1 - I_3} \right]^{1/2} \frac{\pi}{2 F(\pi/2, k^2)}
\hatJ
\ee
when $q< I_1/I_2$ and 
\be
\bar{\bGam}_{\rm rad}^{\rm iso}(q,\pm) = 
\pm \left( \bGam_{{\rm rad}, 0}^{\rm iso} \cdot \ahat_3 \right)
\left[ \frac{I_3 (q-1)}{I_1 -I_3} \right]^{1/2} \frac{\pi}{2 F(\pi/2, k^{-2})}
\hatJ
\ee
when $q > I_1/I_2$.  

Also, 
\be
\overline{\bGam_{\rm rad}^{\rm iso} \cdot \bomega} (q,\pm) \sim \pm
\frac{1}{2} (1 - \gamma_{\rm rad}) u_{\rm rad} a^2 \tilde{\lambda}
\tilde{Q}_{\Gamma}^{\rm iso} \frac{q J}{I_1}~~~.
\ee
Since $(1 - \gamma_{\rm rad}) \tilde{Q}_{\Gamma}^{\rm iso}$ is 
generally less than $\gamma_{\rm rad} |H|$, we can neglect
$\overline{\bGam_{\rm rad}^{\rm iso} \cdot \bomega}$.

\subsubsection{Evaluation of Radiative Torques}

We use the discrete dipole approximation\footnote{We use the code 
	DDSCAT.5a10, which is available at 
	\url{http://astro.Princeton.EDU/$\sim$draine/}.} 
to compute $\bQ_{\Gamma}(\Theta, \Phi, \beta)$ (see Paper I).  
In this exploratory paper, we consider only shape 1 from Paper I and 
$a=0.2 \micron$.  We adopt the dielectric functions for astronomical 
silicate from Li \& Draine (2001).  In order to further reduce the 
computational demands, we assume a monochromatic spectrum, with 
$\lambda = 1.2 \micron$ ($\tilde{\lambda}$ for the ISRF).  In a
future paper, we will consider a range of grain shapes and
sizes and we will fully integrate over the ISRF.  

Computations are only performed for $\Phi =0$, since 
\be
\bQ_{\Gamma}(\Theta, \Phi, \beta) = \bQ_{\Gamma}(\Theta, 0, \beta) \cdot
\ehat_1 \ehat_1 + \bQ_{\Gamma}(\Theta, 0, \beta) \cdot \ehat_2
(\ehat_2 \cos\Phi + \ehat_3 \sin\Phi) + 
\bQ_{\Gamma}(\Theta, 0, \beta) \cdot \ehat_3 (\ehat_3 \cos\Phi - 
\ehat_2 \sin\Phi)~~~.
\ee
We ran the code for 33 values of $\Theta$ between 0 and $\pi$ 
(evenly spaced in $\cos\Theta$) and 32 values of $\beta$ evenly spaced 
between 0 and $2 \pi$.  

\begin{figure}
\epsscale{1.00}
\plotone{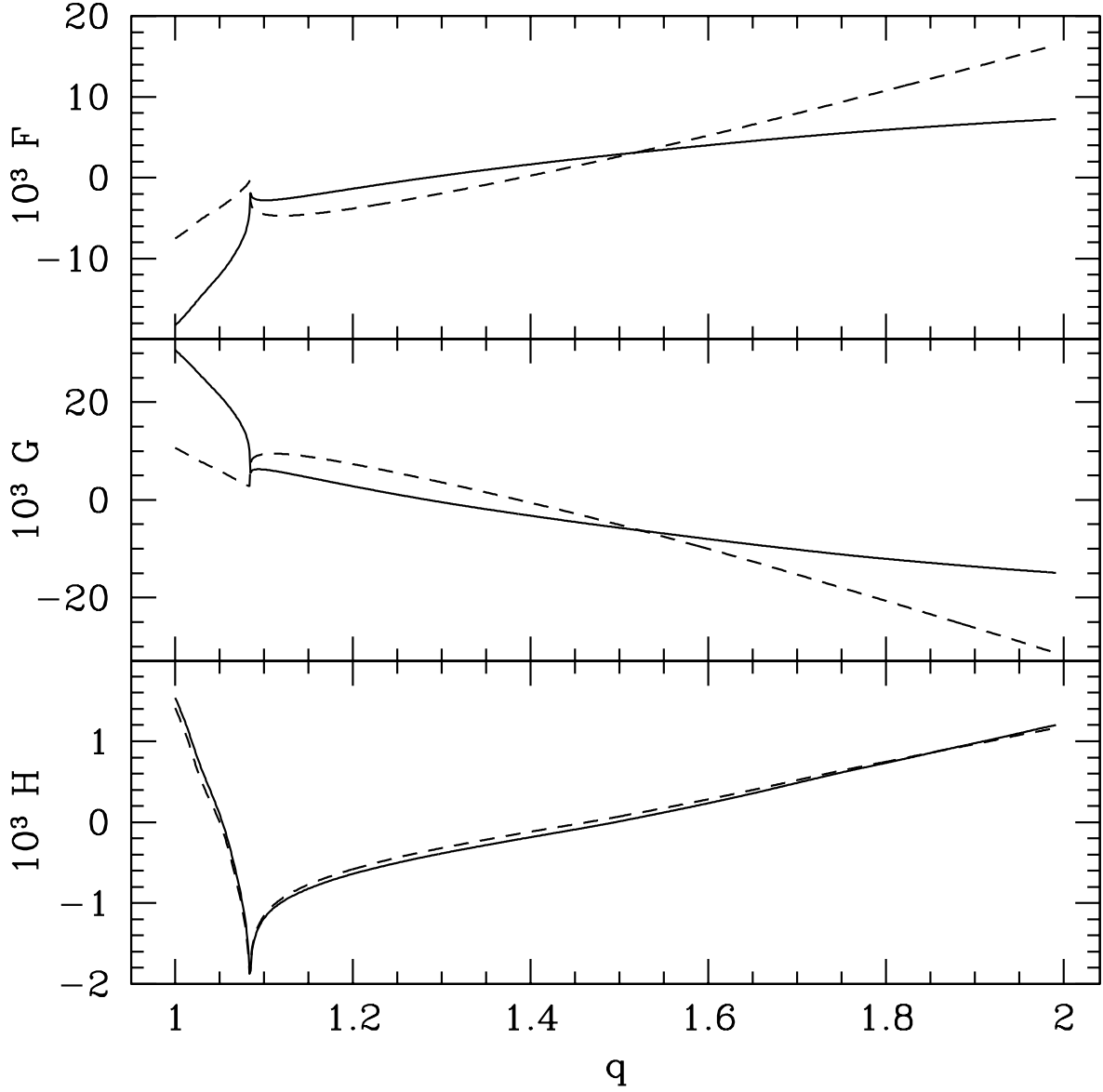}
\caption{
\label{fig:FGH}
$F$, $G$, and $H$ for shape 1, astronomical silicate, $a=0.2 \micron$, 
$\psi = 70\arcdeg$, $\xi=30\arcdeg$, $\phi=160\arcdeg$, and positive (solid) 
and negative
(dashed) flip states.  A monochromatic radiation field with $\lambda =
\tilde{\lambda}_{\rm ISRF} = 1.2 \micron$ was adopted.
        }
\end{figure}

In Figure \ref{fig:FGH}, we plot $F$, $G$, and $H$ versus $q$ for the case 
that $\psi=70\arcdeg$, $\xi=30\arcdeg$, and $\phi=160\arcdeg$.  
For isotropic radiation,
$\bQ_{\Gamma}^{\rm iso} \cdot \ahat_1 = 1.5 \times 10^{-5}$,
$\bQ_{\Gamma}^{\rm iso} \cdot \ahat_2 = 1.2 \times 10^{-6}$, and
$\bQ_{\Gamma}^{\rm iso} \cdot \ahat_3 = -2.2 \times 10^{-7}$.

\section{\label{sec:q_avg} Thermal Averages over $q$}

Since the Barnett dissipation timescale $\tau_{\rm Bar}$ is much shorter
than any other relevant timescales (except for the timescale characterizing
the torque-free motion), we do not follow the evolution of $q$.  Instead,
we average over a thermal distribution of $q$-values when evaluating 
the torques.  We consider the following two cases (see \S 
\ref{sec:strategy_just_flips} for the motivation of these choices). 
In the first case, we assume that the grain is in the positive flip state
with respect to $\ahat_1$ when $q < I_1/I_2$.  When $q \ge I_1/I_2$, we
average over both flip states (with respect to $\ahat_2$ when 
$q=I_1/I_2$ and with respect to $\ahat_3$ when $q > I_1/I_2$).  A quantity 
$A$ averaged in this way is denoted by $\langle A \rangle_+$.  (We assume
that $A$ has already been averaged over torque-free motion and do not 
include a bar to indicate this averaging.)  The second case is identical 
to the first, except that the grain is in the negative flip state with 
respect to $\ahat_1$ when $q < I_1/I_2$; such averaging is denoted by
$\langle A \rangle_-$.  

\subsection{\label{sec:density_of_states} Density of States}

In order to take thermal averages over $q$, we need to evaluate the density 
of states in $q$ (for a grain with constant $J$).  To this end, we
examine the grain's trajectories in phase space.
In general, the phase space for a rotating grain has six dimensions:  one for
each Euler angle plus one for each conjugate momentum.  However, when the
angular momentum is specified this reduces to just one angle plus its 
momentum.  

The Lagrangian $L$ is just the kinetic energy:
\be
L = \frac{1}{2} \sum_i I_i \omega_i^2 = \frac{J^2 \cos^2 \gamma}{2I_1} +
\frac{J^2 \sin^2 \alpha \sin^2 \gamma}{2I_2} + \frac{J^2 \cos^2 \alpha
\sin^2 \gamma}{2I_3}
\ee
(see eqs.~\ref{eq:axis_motion_1} through \ref{eq:axis_motion_3}).  
The $\omega_i$ are related to the Eulerian angles as follows:
\be
\omega_1 = \dot{\zeta} \cos \gamma + \dot{\alpha}~~~,
\ee
\be
\omega_2 = \dot{\zeta} \sin\alpha \sin\gamma + \dot{\gamma} \cos\alpha~~~,
\ee
and
\be
\omega_3 = \dot{\zeta} \cos\alpha \sin\gamma - \dot{\gamma} \sin\alpha~~~.
\ee
Using these relations, we rewrite the Langrangian as
\be
\label{eq:Lagrangian}
L = \frac{J^2}{2I_2 I_3} \left[ I_3 + (I_2 - I_3) \cos^2 \alpha \right] - 
\frac{I_1 I_2 I_3 \dot{\alpha}^2}{2 \left[(I_1 - I_2) I_3 + I_1 (I_2 - I_3)
\cos^2 \alpha \right]}~~~.
\ee

\begin{figure}
\epsscale{1.00}
\plotone{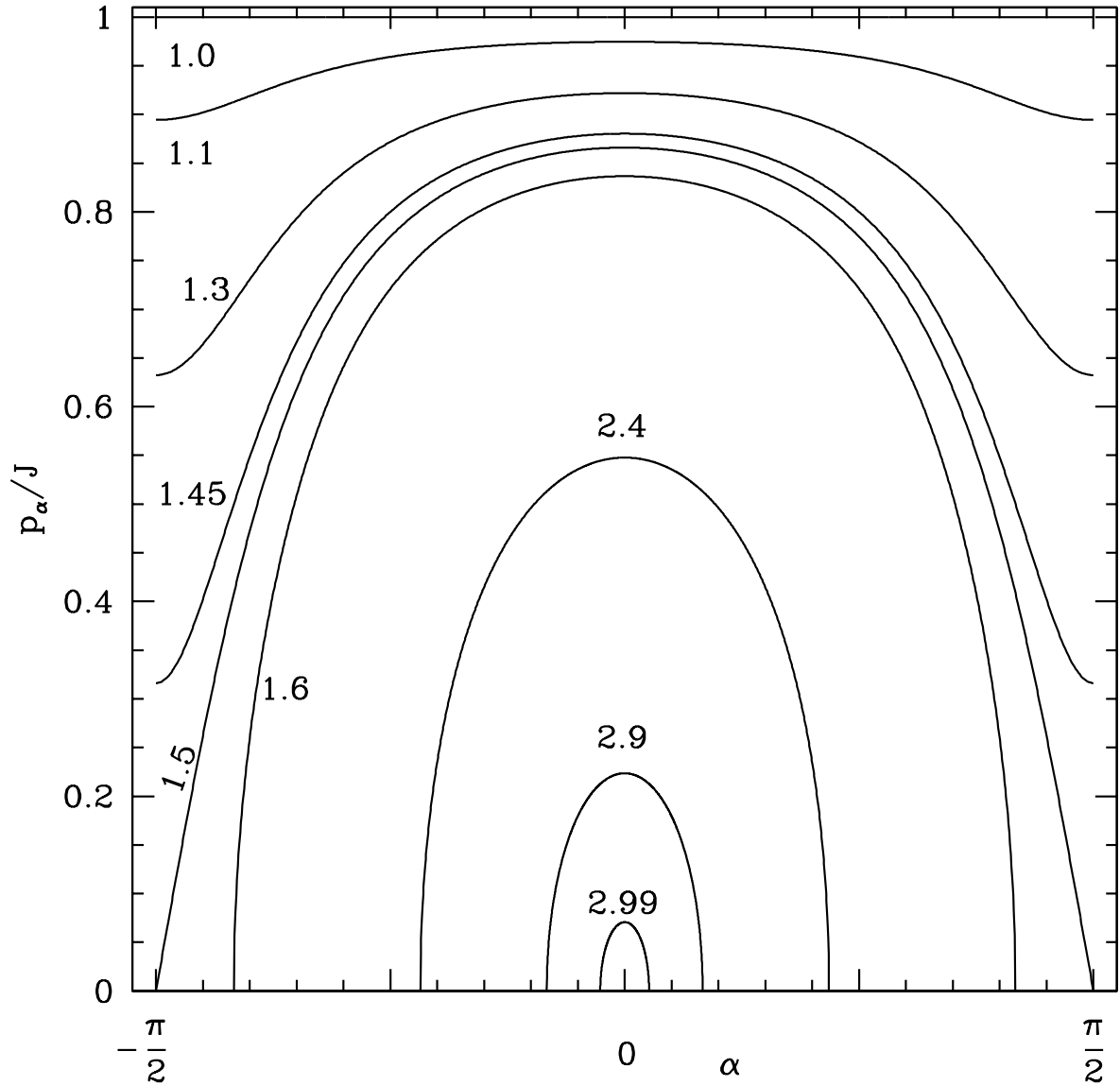}
\caption{
\label{fig:phase_space_traj}
Phase space trajectories for the case that $I_1:I_2:I_3 = 3:2:1$.  The
Eulerian angle $\alpha$ was introduced in \S \ref{sec:axis_motion} and
$p_{\alpha}$ is its conjugate momentum.  The values of $q$ are indicated.
        }
\end{figure}

The conjugate momentum is given by
\be
p_{\alpha} = \frac{\partial L}{\partial \dot{\alpha}} = 
\frac{-I_1 I_2 I_3 \dot{\alpha}}
{\left[ (I_1 - I_2) I_3 + I_1 (I_2 - I_3) \cos^2 \alpha \right]}~~~.
\ee
Expressing the second term in equation (\ref{eq:Lagrangian}) for the 
Lagrangian in terms of $p_{\alpha}$ and setting $L$ equal to the energy
$E$, we find
\be
p_{\alpha} = J \left[ \frac{I_3 (I_1 - I_2 q) + I_1 (I_2 - I_3) \cos^2
\alpha}{I_3 (I_1 - I_2) + I_1 (I_2 - I_3) \cos^2 \alpha} \right]^{1/2}~~~.
\ee
Sample phase space trajectories are plotted in Figure 
\ref{fig:phase_space_traj} for the case that $I_1 : I_2 : I_3 = 3:2:1$.

The number of states with energy between $E_1$ and $E_2$ is proportional 
to the area in phase space bounded by the trajectories for $E_1$ and
$E_2$.  Thus, the number of states with $q$-value between 1 and $q$ is
proportional to the following quantity:
\be
s \equiv 1 - \frac{2}{\pi} \int_0^{\alpha_1} d\alpha \left[
\frac{I_3 (I_1 - I_2 q) + I_1 (I_2 - I_3) \cos^2
\alpha}{I_3 (I_1 - I_2) + I_1 (I_2 - I_3) \cos^2 \alpha} \right]^{1/2}~~~,
\ee
where 
\be
\alpha_1 \equiv \cases{\pi/2 &, $q \le I_1/I_2$\cr
\cos^{-1} \left[ \frac{I_3 (I_2 q -I_1)}{I_1 (I_2 - I_3)} \right]^{1/2}
&, $q > I_1/I_2$\cr}~~~.
\ee
Note that $s$ ranges from 0 to 1 as $q$ ranges from 1 to $I_1/I_3$.

\begin{figure}
\epsscale{1.00}
\plotone{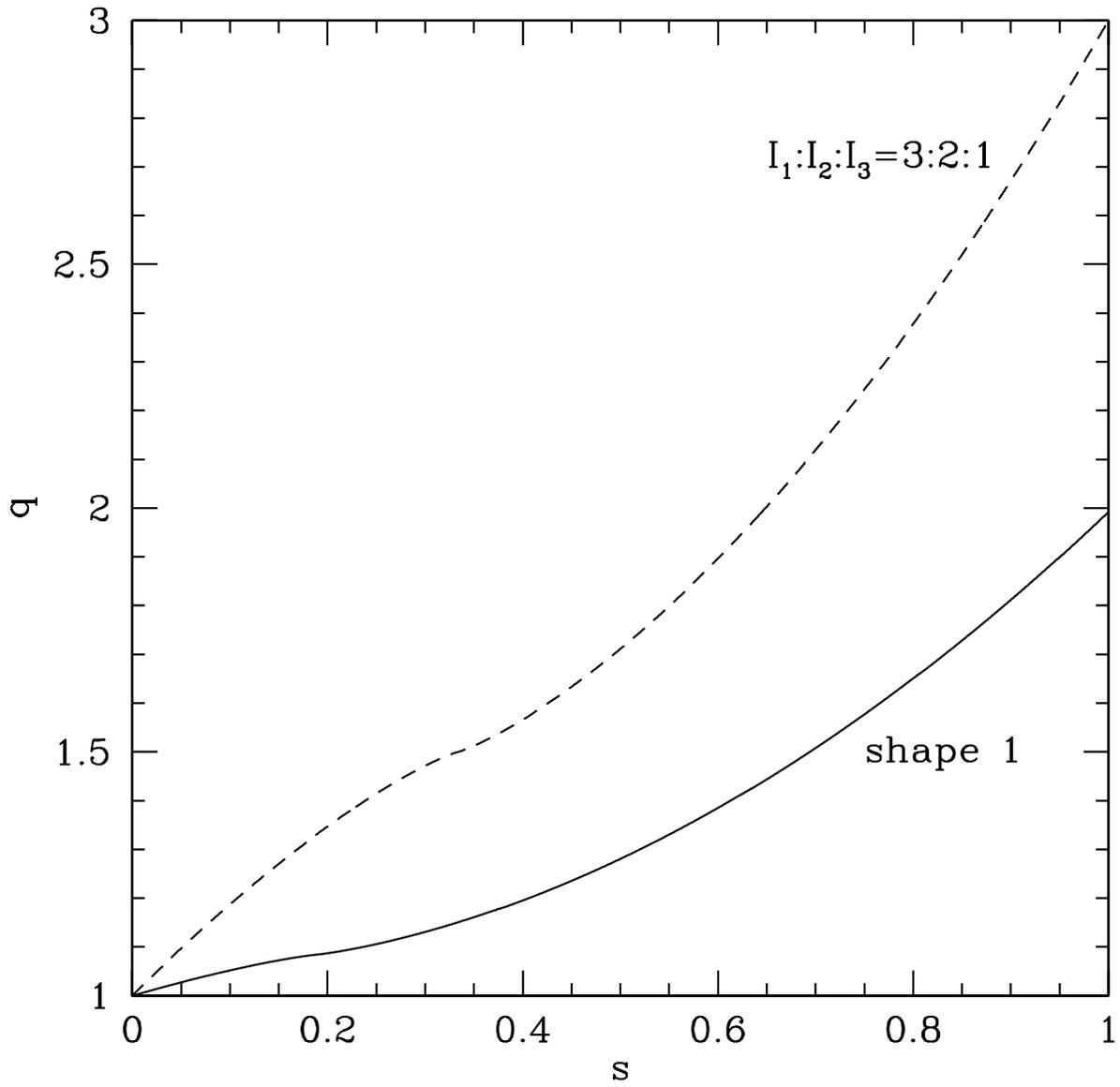}
\caption{
\label{fig:q_s}
$q$ versus $s$ for a grain with $I_1:I_2:I_3 = 3:2:1$
and for shape 1 from Paper I.
        }
\end{figure}

Since the density of states in the variable $s$ is constant, we will 
evaluate thermal averages by integrating over $s$.  
Figure \ref{fig:q_s} shows $q$ as a function of $s$ for a grain with 
$I_1:I_2:I_3 = 3:2:1$ and for shape 1 from Paper I.  

\subsection{\label{sec:avg_torques} Thermal Averages for the Torques}

The average of a function $\bar{A}(q,\pm)$ over a thermal distribution 
of $q$ at constant $J$ is given by
\be
\label{eq:thermal_avg_q}
\langle A \rangle_{\pm} = \frac{\int_0^{s_c} ds \bar{A}[q(s),\pm] 
\exp[-q(s) J^2 /2 I_1 k T_d] + \int_{s_c}^1 ds \, 0.5 \left\{ \bar{A}[q(s),+]
+ \bar{A}[q(s),-] \right\} \exp[-q(s) J^2 /2 I_1 k T_d]}
{\int ds \exp[-q(s) J^2 /2 I_1 k T_d]}~~~,
\ee
where the value of $s$ when $q=I_1/I_2$ is given by
\be
s_c = 1 - \frac{2}{\pi} \sin^{-1} \left[ \frac{I_1 (I_2 - I_3)}
{I_2 (I_1 - I_3)} \right]^{1/2}~~~.
\ee
If $\bar{A}(q)$ does not depend on the flip state, then we will simply 
denote the thermal average over $q$ by $\langle A \rangle$.  

\begin{figure}
\epsscale{1.00}
\plotone{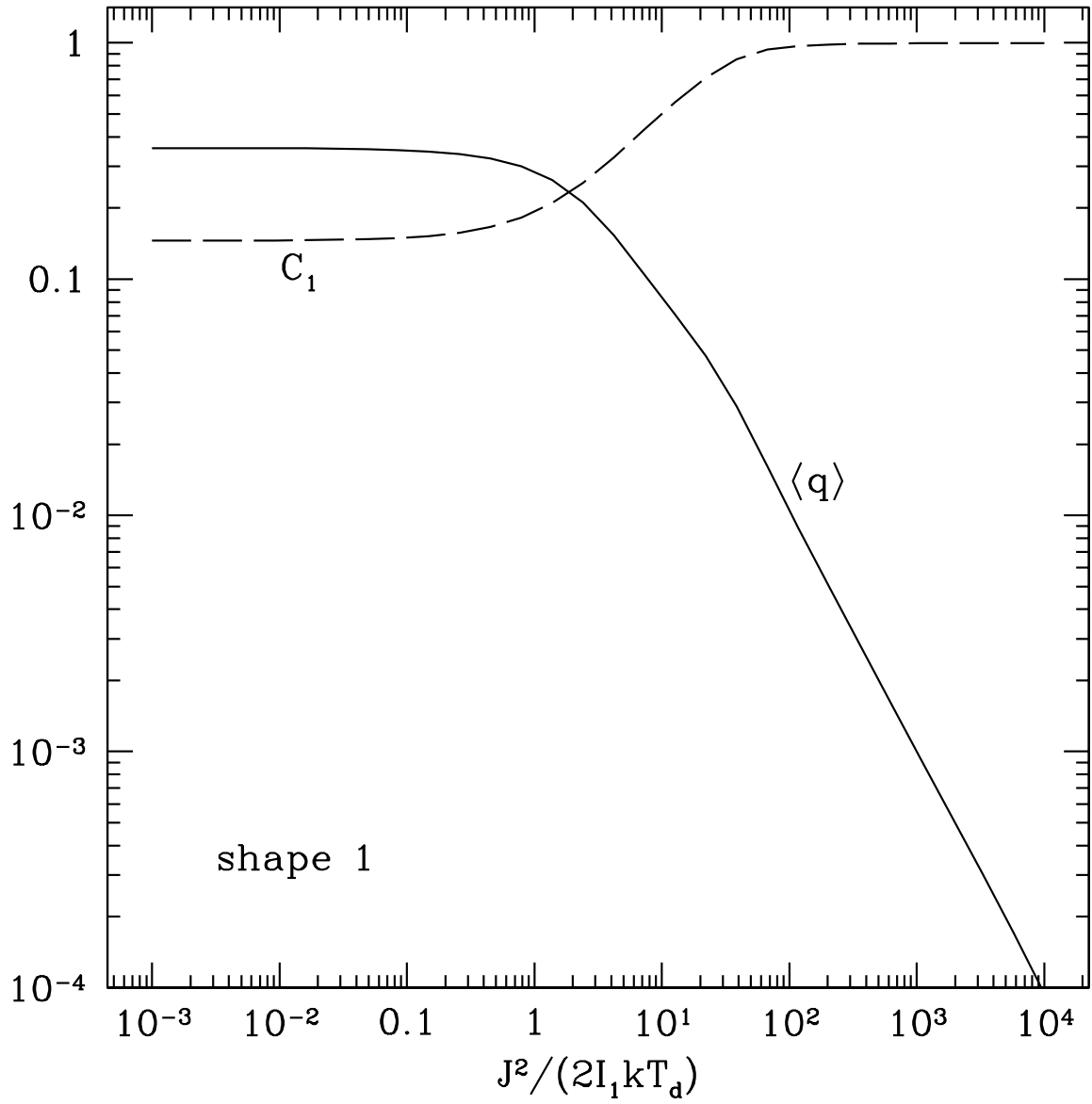}
\caption{
\label{fig:q_q}
$\langle q \rangle$ and $C_1$. 
        }
\end{figure}

The Barnett, drag, and Davis-Greenstein torques are all invariant under
a change in flip state.  With our simple approximations, the Davis-Greenstein
torque does not even depend on $q$, so $\langle \bGam_{\rm DG} \rangle$ is 
given by equation (\ref{eq:D-G_torque}).  We also have approximated the 
Barnett and drag torques as simply proportional to $q$.  Thus, 
$\langle \bGam_{\rm B} \rangle$ is given by equation (\ref{eq:Gamma_Bar_ast}) 
with $q$ replaced by $\langle q \rangle$ and $\langle \bGam_{\rm drag} \rangle
= - \langle q \rangle \bJ / \tau_{\rm drag}$.  In Figure \ref{fig:q_q} we 
plot $\langle q \rangle$ versus $J^2/(2 I_1 k T_d)$.  As 
$J^2/(2 I_1 k T_d) \rightarrow \infty$, $\langle q \rangle \rightarrow 1$ and 
as $J^2/(2 I_1 k T_d) \rightarrow 0$, $\langle q \rangle \rightarrow 
\int_0^1 ds \, q(s)$.  

The H$_2$ formation torque does depend on flip state and its thermal 
average is given by
\be
\langle \bGam_{{\rm H}_2} \rangle_{\pm} = \pm \frac{I_1 \omega_{{\rm H}_2}}
{\tau_{\rm drag, gas}} \left( {\bf \hat{\Gamma}_{{\rm H}_2}} \cdot
\ahat_1 \right) C_1 \hatJ~~~,
\ee
where $C_1 = \langle C \rangle$ and 
\be
\label{eq:C_1}
C = \cases{\left[ \frac{I_1 - I_3 q}{I_1 - I_3} \right]^{1/2}
\frac{\pi}{2 F(\pi/2, k^2)} &, $q < I_1/I_2$\cr
0 &, $q \ge I_1/I_2$\cr}~~~.
\ee
Note that there is no contribution to $\langle \bGam_{{\rm H}_2} \rangle_{\pm}$
from $q \ge I_1/I_2$, since we average over flip states for these $q$, with 
perfect cancellation.  In Figure 
\ref{fig:q_q}, we plot $C_1$ versus $J^2/(2 I_1 k T_d)$; $C_1 \rightarrow 1$
as $J^2/(2 I_1 k T_d) \rightarrow \infty$.   

The thermal average of the torque due to the isotropic component of the 
radiation field is similar to that for the H$_2$ formation torque:
\be
\langle \bGam_{\rm rad}^{\rm iso} \rangle_{\pm} = 
\pm \frac{1}{2} (1 - \gamma_{\rm rad}) u_{\rm rad} a^2 \tilde{\lambda} \,  
\tilde{\bQ}_{\Gamma}^{\rm iso} \cdot \ahat_1 \, C_1 \hatJ~~~.
\ee

\begin{figure}
\epsscale{1.00}
\plotone{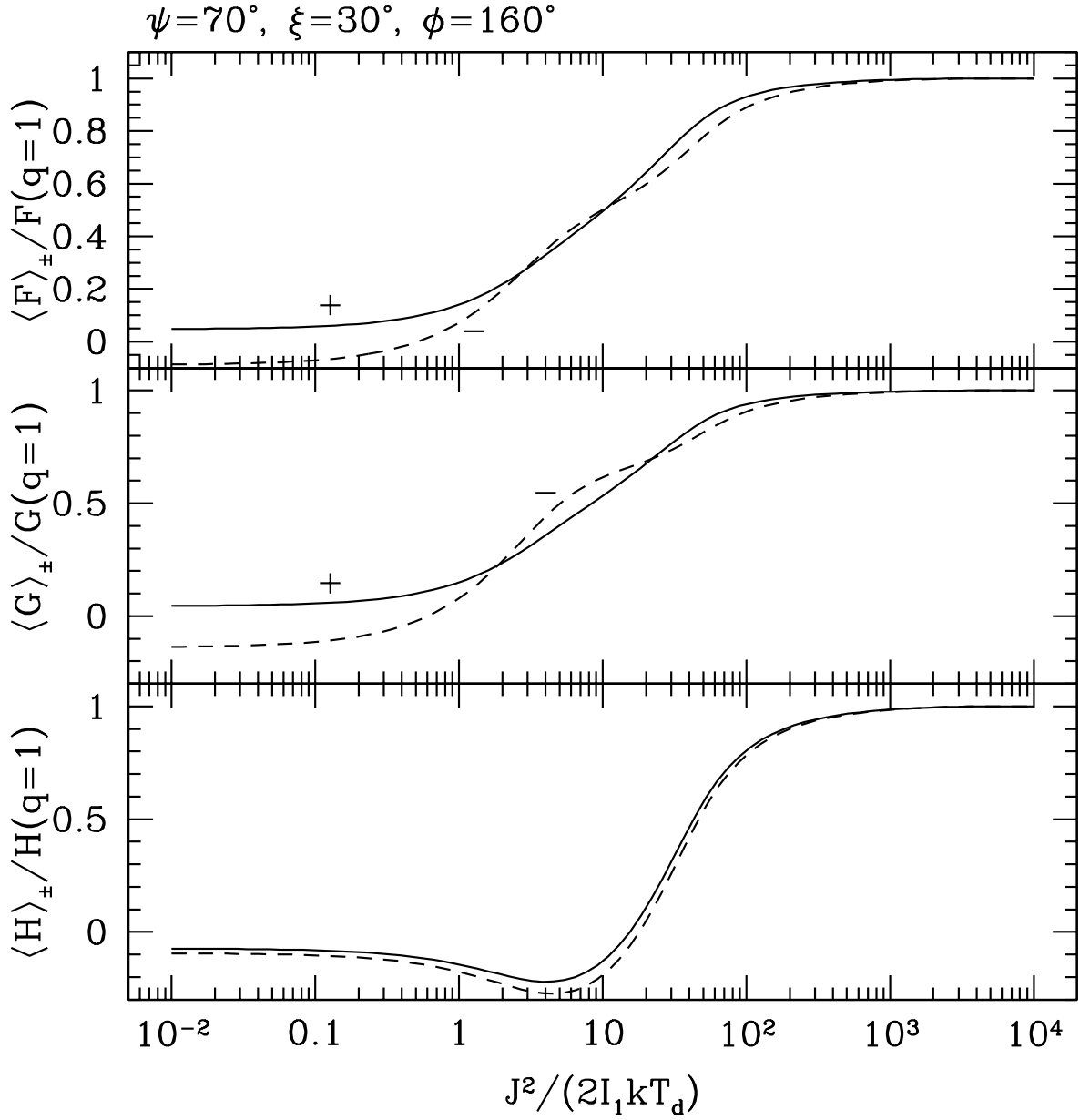}
\caption{
\label{fig:FGH_q}
$\langle F \rangle_{\pm}$, $\langle G \rangle_{\pm}$, and 
$\langle H \rangle_{\pm}$, normalized to their values when $q=1$, for
shape 1, one 
particular angular momentum direction $(\xi,\phi)$ and starlight anisotropy
direction $\psi$.
}
\end{figure}

The thermal average of the torque due to the unidirectional component of 
the radiation field requires the thermal averages $\langle F \rangle_{\pm}$, 
$\langle G \rangle_{\pm}$, and $\langle H \rangle_{\pm}$.  
It is clear from Figure \ref{fig:FGH} that these do not go to zero
when averaging over flip states.  (It is not always the case, however, that  
$H(q)$ for the two flip states are nearly identical.)  In Figure
\ref{fig:FGH_q} we plot $\langle F \rangle_{\pm}$, $\langle G \rangle_{\pm}$, 
and $\langle H \rangle_{\pm}$ versus $ J^2/(2 I_1 k T_d)$
for the case that $\psi = 70\arcdeg$, $\xi=30\arcdeg$, and $\phi = 
160\arcdeg$.  

\section{Equations of Motion \label{sec:eq_of_motion}}

Suppose that during some time interval $\Delta t$ [during which the
grain has angular momentum $\bJ$ defined by $(J,\xi,\phi)$], if the grain
is in a flip state with respect to $\ahat_1$, then 
the probability that it is in the positive flip state
is $f_+$, and the probability that it is in the negative flip state
is $f_-=1-f_+$.  In this case, the 
thermal average of a quantity $A$ is given by 
\be
\langle A \rangle = \langle A \rangle_+ f_+ + \langle A \rangle_- f_-~~~.
\ee
Prescriptions for $f_+$ and $f_-$ are developed in \S 
\ref{sec:strategy_just_flips}.  

The following equations of motion result from equation (\ref{eq:dyn1}):
\be
\label{eq:eq_of_motion_phi}
\frac{d\phi}{dt} = \langle q \rangle \Omega_{\rm B} + 
\frac{\gamma_{\rm rad} u_{\rm rad} a^2 \tilde{\lambda} 
\langle G \rangle(\xi, \phi, \psi, J)}{2 J \sin \xi}~~~,
\ee
\be
\label{eq:eq_of_motion_xi}
\frac{d\xi}{dt} = \frac{\gamma_{\rm rad} u_{\rm rad} a^2 \tilde{\lambda} 
\langle F \rangle(\xi, \phi, \psi, J)}{2J} - \frac{\sin \xi \cos \xi}
{\tau_{\rm DG}}~~~,
\ee
and
\begin{eqnarray}
\label{eq:eq_of_motion_J}
\frac{dJ}{dt} &=& \frac{1}{2} \gamma_{\rm rad} u_{\rm rad} a^2 \tilde{\lambda}
\langle H \rangle(\xi, \phi, \psi, J) - 
\frac{J} {\tau_{\rm drag}} \nonumber \\ &&
+ \left[ \frac{I_1 \omega_{{\rm H}_2}}{\tau_{\rm drag, gas}} 
{\bf \hat{\Gamma}_{{\rm H}_2}} + \frac{1}{2} (1-\gamma_{\rm rad})
u_{\rm rad} a^2 \tilde{\lambda} \tilde{\bQ}_{\Gamma}^{\rm iso} \right]
\cdot \ahat_1 C_1 (f_+ - f_-)
- \frac{J \sin^2 \xi}{\tau_{\rm DG}}~~~.
\end{eqnarray}
These equations are similar to equations (24) through (26) in Paper II 
(except note the sign error involving $\Omega_{\rm B}$ in eqs.~10 and 24
in Paper II).  The dependence of $\langle F \rangle$, $\langle G \rangle$, 
and $\langle H \rangle$ 
on $J$ results from the presence of $J$ in equation (\ref{eq:thermal_avg_q}).

\begin{figure}
\epsscale{1.00}
\plotone{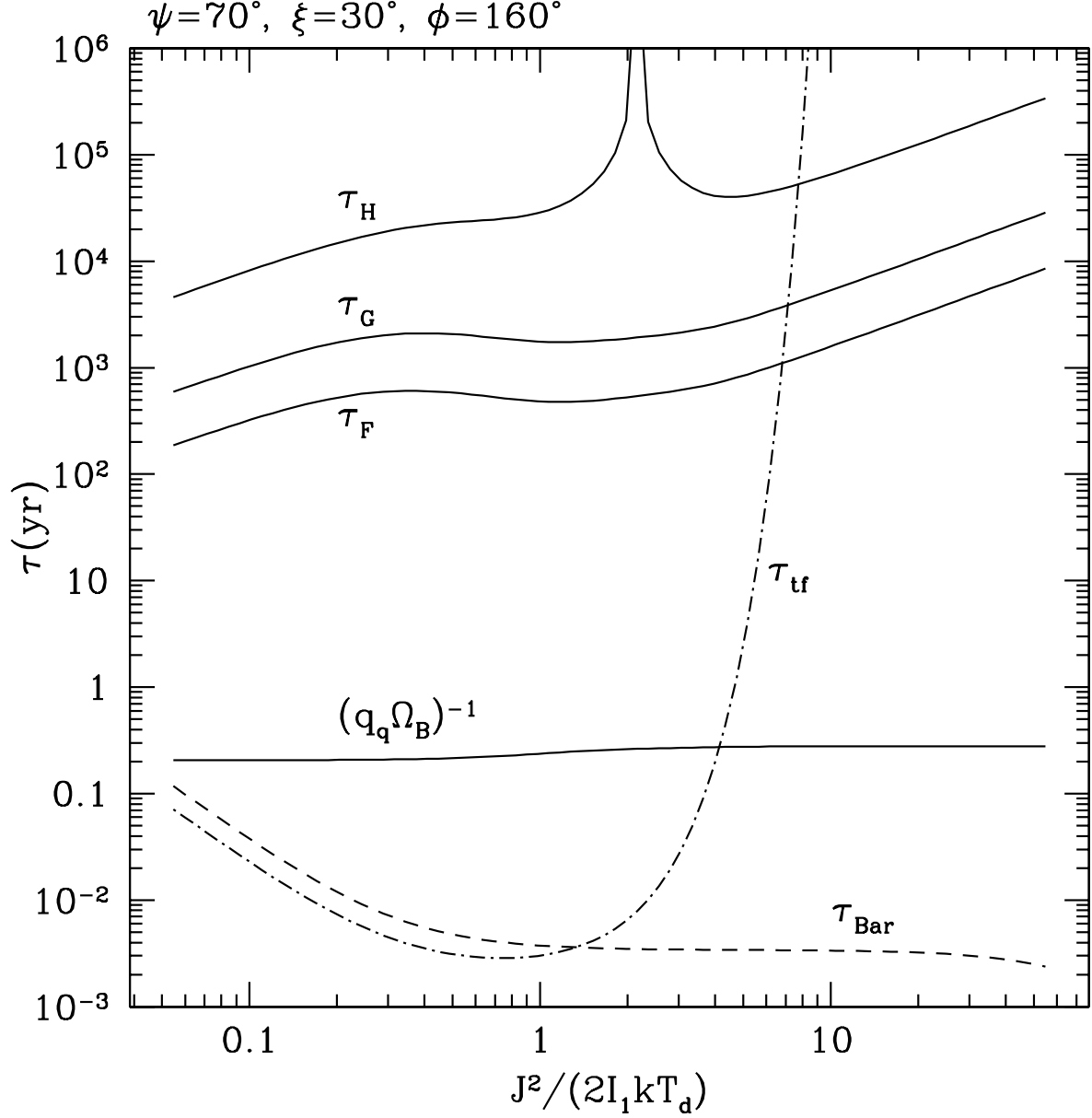}
\caption{
\label{fig:timescales2}
The Barnett dissipation timescale $\tau_{\rm Bar}$, the thermal flipping 
timescale $\tau_{\rm tf}$, and timescales associated with various terms
in the equations of motion:  $\tau_F \equiv 2J/(\gamma_{\rm rad} 
u_{\rm rad} a^2 \tilde{\lambda} \langle F \rangle_+)$, $\tau_G \equiv 
2J \sin\xi/(\gamma_{\rm rad} u_{\rm rad} a^2 \tilde{\lambda} 
\langle G \rangle_+)$, and $\tau_H \equiv 2J/(\gamma_{\rm rad} u_{\rm rad} 
a^2 \tilde{\lambda} \langle H \rangle_+)$, for shape 1,
$a=0.2\micron$,
$T_d = 15\K$, $\psi =70\arcdeg$, $\xi=30\arcdeg$, and $\phi=160\arcdeg$.
	}
\end{figure}

In Figure \ref{fig:timescales2} we plot the dynamical times associated
with the terms involving $\langle F \rangle_+$, $\langle G \rangle_+$, and 
$\langle H \rangle_+$ ($\tau_F$, $\tau_G$, and $\tau_H$, defined in the 
figure caption), as well as the magnetic precession timescale 
$(\langle q \rangle \Omega_{\rm B})^{-1}$, the Barnett
dissipation timescale $\tau_{\rm Bar}$, and the thermal flipping 
timescale $\tau_{\rm tf}$ (see \S \ref{sec:strategy_just_flips}),  
versus $J/(I_1 \omega_{{\rm T},1})$, 
for the case that $T_d=15 \K$, $\psi =70\arcdeg$, $\xi=30\arcdeg$, and 
$\phi=160\arcdeg$.  As mentioned in \S \ref{sec:q_avg}, $\tau_{\rm Bar}$
is much shorter than the other timescales (although note that 
$\tau_F$, $\tau_G$, and $\tau_H$ are shorter when the entire ISRF is 
adopted).  

Since $(\langle q \rangle \Omega_{\rm B})^{-1}$
is much shorter than $\tau_F$, $\tau_G$, and $\tau_H$ for all 
$J^2/(2 I_1 k T_d)$ in Figure \ref{fig:timescales2} (and the 
radiative torques are generally expected to be more important than the 
Davis-Greenstein and H$_2$ formation torques; see Paper II), we expect
that we will generally be able to average the equations of motion over
grain precession.  
In this case, only equations (\ref{eq:eq_of_motion_xi}) and 
(\ref{eq:eq_of_motion_J}) remain, and $\langle F \rangle(\xi, \phi, \psi, J)$ 
and $\langle H \rangle(\xi, \phi, \psi, J)$ are averaged over the motion in 
$\phi$.  Thus, we define
\be
\label{eq:Fphiavgpm}
\Fphiavgpm (\xi, \psi, J) \equiv \frac{1}{\tau_{\phi}}
\int_0^{2 \pi} d\phi \left| \frac{d\phi}{dt}
\right|^{-1} \langle F \rangle_{\pm}(\xi, \phi, \psi, J) 
\ee
and
\be
\label{eq:Hphiavgpm}
\Hphiavgpm (\xi, \psi, J) \equiv \frac{1}{\tau_{\phi}}
\int_0^{2 \pi} d\phi \left| \frac{d\phi}{dt}
\right|^{-1} \langle H \rangle_{\pm}(\xi, \phi, \psi, J)~~~,
\ee
where 
\be
\tau_{\phi} \equiv \int_0^{2 \pi} d\phi |d\phi / dt|^{-1}~~~.
\ee
When $\tau_G \gg (\langle q \rangle \Omega_{\rm B})^{-1}$, $d\phi/dt$ is 
nearly independent of $\phi$ and $\Fphiavgpm$ and $\Hphiavgpm$ have 
the form
adopted in Paper II (these were denoted $\bar{F}$ and $\bar{H}$ in
Paper II).  If $|\tau_G| < (\langle q \rangle \Omega_{\rm B})^{-1}$ 
and $\tau_G(\phi)$ undergoes sign changes, then the grain might not even 
undergo complete cycles of 0 to $2\pi$ in $\phi$.  However, we expect that
this will only occur for $\xi \rightarrow 0$ or $\pi$.

When averaging over the motion in $\phi$, we replace 
$\langle F \rangle(\xi, \phi, \psi, J)$ and 
$\langle H \rangle(\xi, \phi, \psi, J)$ in 
equations (\ref{eq:eq_of_motion_xi}) and (\ref{eq:eq_of_motion_J}) with 
$\Fphiavgplus (\xi, \psi, J) f_+ + \Fphiavgminus (\xi, \psi, J) f_-$ 
and 
$\Hphiavgplus (\xi, \psi, J) f_+ + \Hphiavgminus (\xi, \psi, J) f_-$, 
where the 
evaluation over the thermal averaging cases is done after the averaging
over the motion in $\phi$.  
When $\tau_{\rm tf} < (\langle q\rangle \Omega_B)^{-1}$, 
these evaluations should actually be done in the reverse order, 
but in that case it would not be possible to construct an 
interpolation table for $\Fphiavgpm$ and $\Hphiavgpm$ prior to 
running the evolution code
(see \S \ref{sec:strategy_just_flips}).  Since it is almost always the case
that $\tau_G \gg (\langle q \rangle \Omega_{\rm B})^{-1}$, this reversal of  
order will not introduce significant error.

The behavior of $\Fphiavgpm (\xi, \psi, J)$ in the limit $\xi \rightarrow 0$
or $\pi$ is somewhat subtle; see Appendix \ref{app:sinxi->0} for a detailed
discussion.

\section{\label{sec:strategy_just_flips} Algorithm 
for Evolving Dynamics With Thermal Flips as the Only Stochastic Element}

As mentioned in \S \ref{sec:intro}, we expect stochasticity due to random 
gas atom impacts and H$_2$ formation to be less important than random thermal 
flipping due to Barnett relaxation.  Thus, as a first approximation, we
will evolve the grain dynamics without taking the 
former stochasticity into account.  In a future paper
we will describe a more detailed analysis that accounts for this 
stochasticity in an approximate way.  

In order to ensure a consistent treatment of the stochastic flipping 
throughout the entire dynamical evolution, we integrate the equations 
of motion using a constant time step size $\Delta t$.  
We do not expect that there would be any significant biases if a scheme
with a variable step size were employed.  However, when we generalize 
further (in a future paper) to include stochasticity associated with 
gas atom impacts and H$_2$ formation, we will adopt Langevin equations,
for which constant step sizes are preferred.  Even with a constant step size,
the computer time spent running the evolution code is insignificant compared
with the time devoted to the scattering calculations and construction 
of interpolation tables for $\Fphiavgpm$ and $\Hphiavgpm$.

For sufficiently large $J$, the grain will rarely have $q \ge I_1/I_2$. 
In this limit, we can consider the grain to always be in one of the flip
states with respect to $\ahat_1$, with some probability per unit time
$\tau_{\rm tf}^{-1}$ that a flip will occur.
We adopt the simple approximation suggested by Lazarian \& Draine (1999a):
\be
\label{eq:tau_tf}
\tau_{\rm tf}^{-1} \sim \tau_{\rm Bar}^{-1} \exp \left\{ - \frac{1}{2}
\left[ \left( \frac{J}{J_0} \right)^2 -1 \right] \right\}~~~,
\ee
where
\be
J_0 \equiv \left( \frac{I_1 I_2 k T_d}{I_1 - I_2} \right)^{1/2} =
3.2 \times 10^{-20} \left( \frac{\alpha_1 \alpha_2}{\alpha_1 - \alpha_2}
\right)^{1/2} \left( \frac{\rho}{3 \g \cm^{-3}} \right)^{1/2}
\left( \frac{T_d}{15 \K} \right)^{1/2} \left( \frac{a}{0.1 \micron}
\right)^{5/2} \erg \s~~~.
\ee

During time interval $\Delta t$,
the grain spends some fraction $f_0$ of the time in its original flip state,
and a fraction $(1-f_0)$ in the opposite flip state.
Ideally, we would construct the distribution function for $f_0$ and 
randomly select an $f_0$ from this distribution for each time step. 
However, it is not clear how to construct this distribution function.  
The expectation value $f_{\rm same}(\Delta t)$ 
for the fraction of the time step that the grain 
will be in its original flip state is easily obtained:
\be
\label{eq:f_same}
f_{\rm same} (\Delta t) = \frac{1}{\Delta t} \int_0^{\Delta t} 
P_{\rm same}(t) \, dt = 
\frac{1}{2} \left[ 1 + \frac{\exp(-\Delta t/ \tau_{\rm tf})
\sinh(\Delta t /\tau_{\rm tf})}{\Delta t/ \tau_{\rm tf}}\right]~~~, 
\ee
where 
\be
\label{eq:P_same}
P_{\rm same}(\Delta t)= \exp(-\Delta t/ \tau_{\rm tf}) \cosh(\Delta t / 
\tau_{\rm tf})
\ee
is the probability that the grain is in the same flip state 
after a time $\Delta t$
(see Appendix D for the derivation of this result).  
As a simple approximation, we could set $f_0 = f_{\rm same}$ for each 
time step.  This would represent ``typical'' behavior'', but
would deny the grain the chance of ever spending a 
significant period of time in one particular flip state.  Such a ``run''
in one flip state could occur in reality, and could have the 
important consequence of allowing a grain that would otherwise be stuck 
in the thermal rotation regime to escape to suprathermal rotation.  

Thus, we adopt the following slightly more complicated prescription.  The
probability that the grain undergoes no flips during the time step is
\be
P_0 = \exp(-\Delta t / \tau_{\rm tf})~~~.
\ee
We take the grain to remain in its original flip state throughout the 
time step with probability $P_0$.  If the grain does not remain in its
original flip state throughout the time step, then we assume that the grain
spends a fraction $f_s$ of the step in its original flip state, where 
$f_s$ is determined from $P_0 + (1-P_0) f_s = f_{\rm same}$:
\be
\label{eq:f_s}
f_s = \frac{f_{\rm same}-P_0}{1-P_0} ~~~.
\ee
Thus, if the grain is in the positive flip state at
the start of the time step, then $f_+ = 1$ and $f_- = 0$ with probability
$P_0$; otherwise, $f_+ = f_s$ and $f_- = 1-f_s$.  

The flip state at the start of each time step is a stochastic variable.  As 
mentioned above, we take the grain to remain in its original flip state 
throughout the entire step with probability $P_0$.  In this case, the 
flip state at the start of the next step is taken to be the same as it is
in the current step.  Otherwise, the flip state at the start of the next
step is taken to be the opposite of the flip state at the start of the 
current step with probability $\Pf$, 
determined from $1 - P_{\rm same} = (1-P_0) \Pf$:
\be
\Pf = \frac{1 - P_{\rm same}}{1-P_0}~~~.
\ee
Table \ref{tab:branches} 
summarizes the 3 branches which can be taken on each time step.

In the limit of low $J$, the grain does spend a significant fraction of
the time with $q \ge I_1/I_2$.  Furthermore, the grain undergoes rapid
flipping in this limit, so that the grain spends half of its time with 
$q > I_1/I_2$ in the positive flip state with respect to $\ahat_3$ and
and half of the time in the negative flip state.  This motivates the 
definition of our two thermal averaging cases, given at the beginning of 
\S \ref{sec:q_avg}.  

If $\Delta t > \tau_{\phi}$, then we average over motion in $\phi$; 
otherwise we do not.  If $\phi$-averaging mode applied in the previous 
time step but does not apply in the current step, then a value of 
$\phi$ is selected randomly at the start of the current time step.  
Since we expect that $\phi$-averaging mode
will always apply, we tabulate 
$\langle F \rangle_{\pm}[\cos\xi, J^2/(2I_1 kT_d)]$, 
$\langle H \rangle_{\pm}[\cos\xi, J^2/(2I_1 kT_d)]$, and 
$\tau_{\phi}[\cos\xi, J^2/(2I_1 kT_d)]$ 
for a given $\psi$ before running the dynamical evolution code and 
interpolate as needed.  If 
non-$\phi$-averaged quantities are needed at any point, then they are 
calculated within the evolution code.  

\section{\label{sec:results} Results}

As an example of our 
new method for evolving the grain dynamics including the effects of
thermal fluctuations and thermal flipping,
we consider shape 1, $a=0.2 \micron$, a monochromatic radiation field
with $\lambda = \tilde{\lambda}_{\rm ISRF} = 1.2 \micron$, $u_{\rm rad} = 
u_{\rm rad, ISRF}$, $\gamma_{\rm rad} = 0.1$, $T=100\K$, $\nH=30 \cm^{-3}$,
$T_d = 15\K$, $B=5 \mu G$, $\rho = 3 \g \cm^{-3}$, 
and $\psi = 70\arcdeg$.  Because we
adopt a monochromatic radiation field with $\lambda=1.2\micron$ 
(as opposed to using a realistic spectrum for interstellar starlight),
this example will
be illustrative of a case where the radiative torques---while still dynamically
very important---are relatively weak (corresponding
to $a\approx0.1\micron$ grains with a realistic starlight spectrum).

For simplicity, we
ignore the torques due to 
infrared emission (which for this case is negligible compared to gas drag)
and H$_2$ formation.  In future 
work we will ascertain the importance of the H$_2$ formation torque.

\begin{figure}
\epsscale{1.00}
\plotone{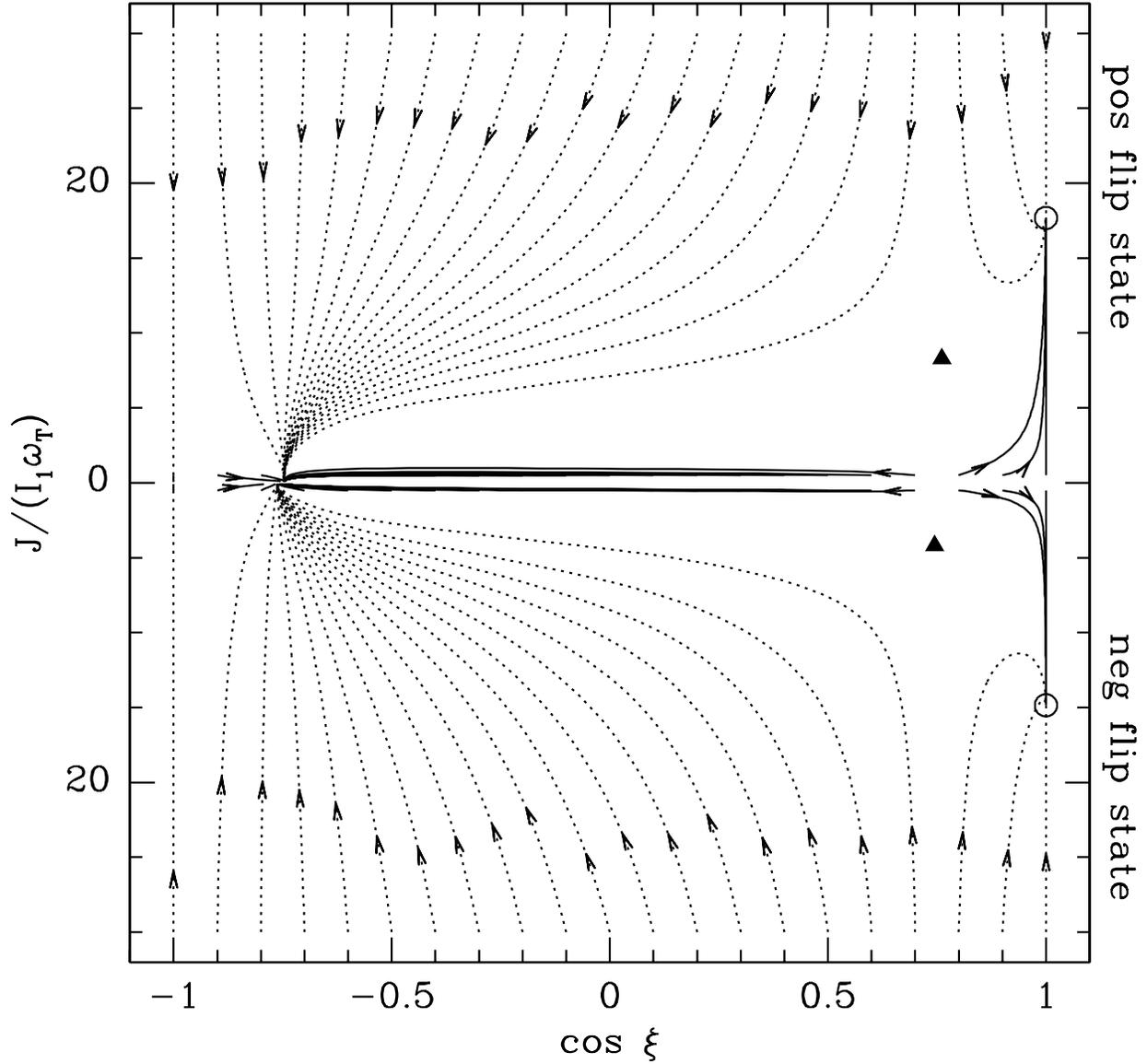}
\caption{
\label{fig:suprathermal_psi70}
Trajectories computed using the method of Paper II (i.e., constraining 
$\bJ$ to be parallel or anti-parallel to $\ahat_1$) 
for $\psi=70\arcdeg$ 
(see text for specification of other 
parameter values).  Dotted (solid) curves originate at 
$J/I_1 \omega_{\rm T} =30$ (0.5).  Attractors (repellors) are indicated
by open circles (filled triangles).  
        }
\end{figure}

In Figure \ref{fig:suprathermal_psi70}, we plot the trajectory map 
calculated
using the method of Paper II, in which it is assumed that $\bJ$ always lies
along $\ahat_1$ (parallel or anti-parallel).    
The dotted (solid) trajectories originate at 
$J/I_1 \omega_{\rm T}= 30$ (0.5) and the upper (lower) half-plane
is for the positive (negative) flip state.  

If we had included this map in Paper II, then it would have looked quite
different.  In Paper II, $\xi$ was defined as the polar angle of 
$\ahat_1$ in alignment coordinates, whereas here we have defined $\xi$
as the polar angle of $\bJ$ in alignment coordinates.  Henceforward, we
will write $\xi_{\rm II}$ for the angle as defined in Paper II.  
When the grain is in the positive flip state with 
respect to $\ahat_1$, $\xi_{\rm II} = \xi$, but in the negative flip 
state, $\xi_{\rm II} = \pi - \xi$.  In Paper II, negative flip states 
were indicated by assigning a negative value to $J$; here $J \equiv |\bJ|$, 
which is always positive.  Since trajectory maps in Paper II were drawn 
in terms of $\cos \xi_{\rm II}$, the lower half of the map in this paper
(i.e., below the line $J=0$) must be reflected in the line $\cos\xi=0$ in 
order to look like the maps in Paper II (and vice versa).  

Note the near symmetry of the map under reflection in the line $J=0$.  
The maps in Paper II also appear somewhat symmetric, though to a lesser 
degree than here.  (In the maps in Paper II, the symmetry is with respect to 
a $180\arcdeg$ rotation.) 
Evidently the radiative torque changes less under a
grain flip for a monochromatic radiation spectrum with $\lambda =1.2 \micron$
than for the ISRF (at least for shape 1).

From Figure \ref{fig:suprathermal_psi70}, we see that there are two 
stable stationary points (``attractors''), with $\cos\xi =1$ and 
$J/I_1 \omega_{\rm T}=17.7$ (14.8) for the positive (negative) flip 
state.  There are also two unstable stationary points (``repellors''), 
with $(J/I_1 \omega_{\rm T}, \cos \xi) = (8.3, 0.76)$ for the positive
flip state and $(J/I_1 \omega_{\rm T}, \cos \xi) = (4.2, 0.74)$ for
the negative flip state.

Crossover points, where the trajectories cross $J=0$, are located at 
$\cos \xi_{\rm II} = -1$, -0.75, 0.77, and 1.  (Those at 
$\cos \xi_{\rm II} = -1$ and 1 are ``crossover repellors''.  Only one 
trajectory passes through each of these crossovers; thus, they are not
physically relevant.)  Suppose $\bJ$ really were
constrained to lie along $\ahat_1$, for all $J$.  In this case, a grain 
could enter the crossover at $\cos \xi_{\rm II}=-0.75$ from above, i.e., in
the positive flip state.  It would then emerge in the negative flip state,
still with $\cos \xi_{\rm II}=-0.75$, in which case $\cos \xi = 0.75$.  
The direction of $\ahat_1$ remains unchanged during the grain flip,
while $\bJ$ flips.  In other words, the grain's orientation does not 
change; it stops spinning, and then begins spinning in the opposite sense.
Some of the trajectories emerging from this crossover proceed to the 
crossover at $\cos \xi_{\rm II} = 0.77$ ($\cos \xi = -0.77$) and others lead 
to the attractor at $\cos \xi_{\rm II}=-1$ ($\cos \xi =1$).   
Trajectories emerge from 
the crossover at $\cos \xi_{\rm II} = 0.77$ with the grain in the positive
flip state, with some heading towards the crossover at 
$\cos \xi_{\rm II}=-0.75$ and others heading towards the 
stationary point at $\cos \xi_{\rm II}=1$.  Thus, in the terminology of 
Paper II, this is a ``semi-cyclic'' map.  

The dynamical picture changes in two distinct ways when the 
formalism developed in this paper is adopted.  First, since we no longer 
enforce $\ahat_1 \parallel \bJ$, the structure of the torques as functions
of $J$, $\xi$, and flip state are different.  Thus, even in the absence of
thermal flips, we can expect qualitative changes in the structure of the 
trajectory map at low $J$.  Second, it is no longer possible to construct
strictly deterministic trajectory maps, because of the random thermal flipping.
In order to gauge the importance of these two effects, we will progress
in stages, first suppressing the thermal flips.  

\begin{figure}
\epsscale{1.00}
\plotone{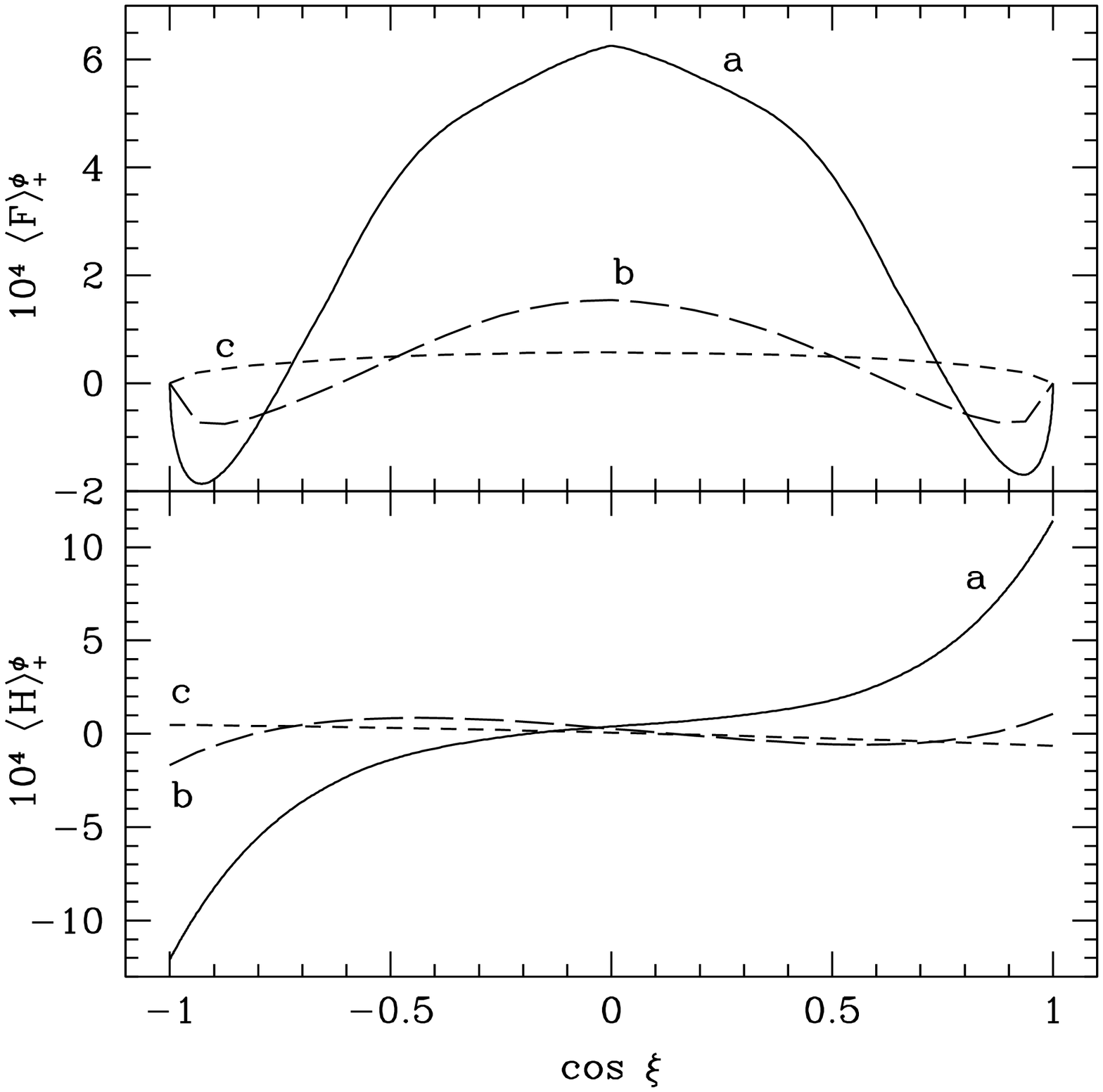}
\caption{
\label{fig:FH_q_bar}
$\Fphiavgplus$ and $\Hphiavgplus$ versus $\cos \xi$ for 
shape 1 and $\psi=70\arcdeg$.
a) $J \rightarrow \infty$; 
b) $J/I_1 \omega_{\rm T} = 1.7$ (for $T_{\rm gas}=100 \K$ and $T_d=15 \K$); 
c) $J \rightarrow 0$.  
        }
\end{figure}

In Figure \ref{fig:FH_q_bar}, we display
$\Fphiavgplus(\xi)$ and $\Hphiavgplus(\xi)$
for $J \rightarrow \infty$, for $J \rightarrow 0$, and for an intermediate
$J$; $\Fphiavgminus(\xi)$ and $\Hphiavgminus(\xi)$ are nearly identical to 
$\Fphiavgplus(\xi)$ and $\Hphiavgplus(\xi)$ at low $J$.
The low-$J$ versions of these functions are quite different from the 
high-$J$ versions.  Since the locations of the stationary points and 
crossovers depend on the structure of these functions, this implies that
the locations of crossovers and low-$J$ stationary points could 
differ from the locations inferred using the analysis of Paper II.  
Indeed, there could be crossovers or stationary points not even indicated
by the analysis of Paper II.  Alternatively, such points found using the 
Paper II formalism might turn out not to exist after all.  In Paper II,
we showed that crossovers occur when $\langle F\rangle^{\phi}(\xi)=0$.  Since 
$\langle F\rangle^{\phi}$ only has zeros at $\cos \xi = \pm 1$ as 
$J \rightarrow 0$, we may expect to no longer
find crossovers at $\cos \xi_{\rm II} = -0.75$ and $0.77$.

\begin{figure}
\epsscale{1.00}
\plotone{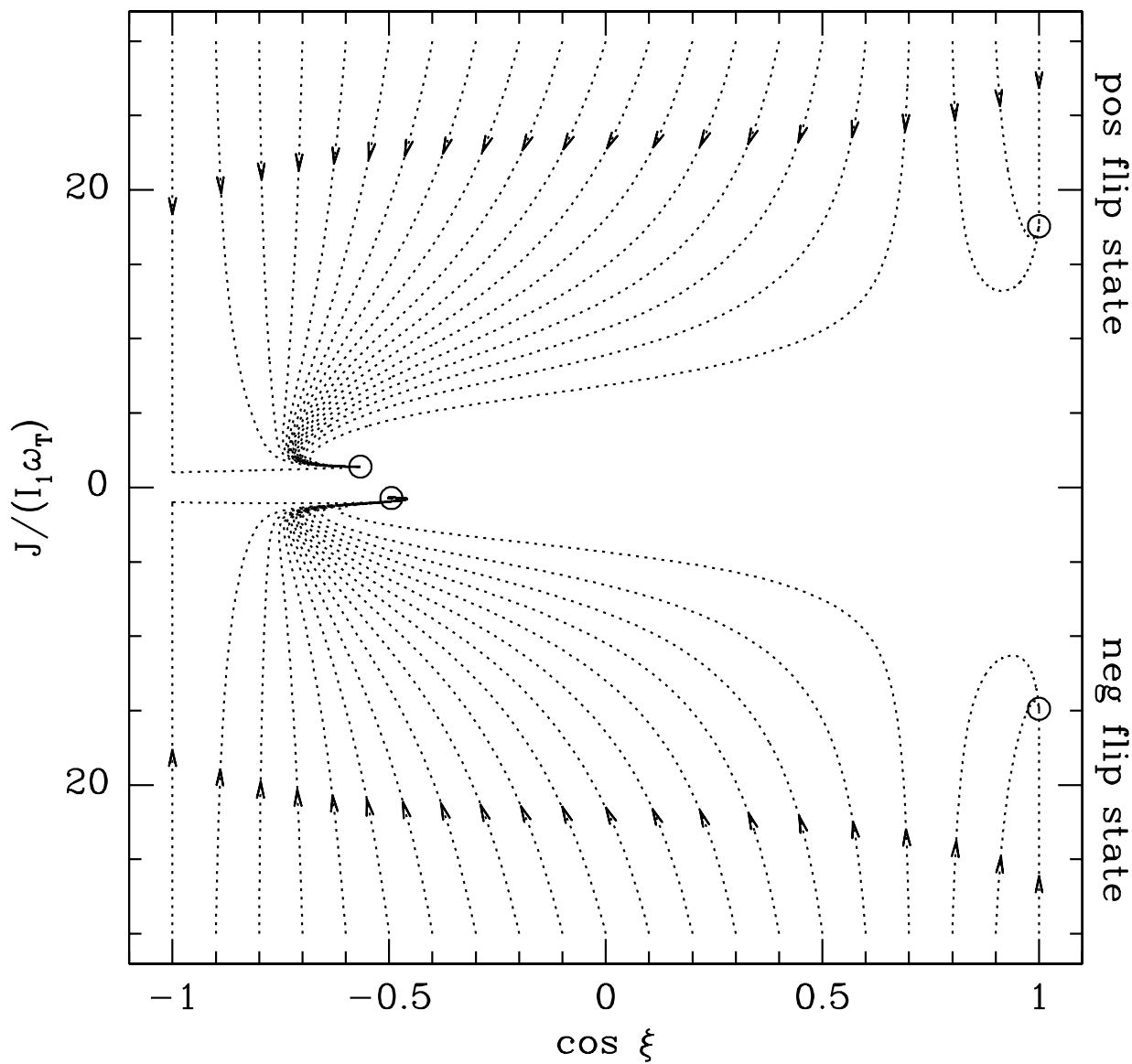}
\caption{
\label{fig:psi70_highJ_noflips}
Trajectory map (with same parameters as for Figure 
\ref{fig:suprathermal_psi70}) computed using the method developed in this
paper (i.e., relaxing the constraint that $\bJ$ must be parallel or 
anti-parallel to $\ahat_1$), except that thermal flipping is prohibited 
(see text for details). 
       }
\end{figure}

Figure \ref{fig:psi70_highJ_noflips} is the trajectory map (with 
$J/I_1 \omega_{\rm T} = 30$ initially) constructed using the method 
developed in this paper, except that thermal flipping is 
prohibited: a grain which begins the trajectory in the positive flip state
has $f_+=1$ and $f_-=0$ throughout (and vice versa for the negative flip 
state).  Since thermal flipping is not included,
these are ``deterministic'' trajectories: they are fully determined
once the initial conditions (including flip state) are specified.
Note the dramatic
change in the map:  two new attractors have appeared, at 
$(J/I_1 \omega_{\rm T}, \cos \xi) = (1.40, -0.57)$ (positive flip state)
and $(0.71, -0.49)$ (negative flip state), and the crossovers have 
disappeared.  The map is now non-cyclic.  

The low-$J$ attractors arise because $\Hphiavgpm$ are smaller in the 
low-$J$ limit than in the high-$J$ limit due to orientational 
averaging---at small $J$, thermal fluctuations allow the grain to spend a 
large
fraction of the time with large angles $\gamma$ between $\bJ$ and $\ahat_1$
(see Fig.~\ref{fig:FH_q_bar}; curves 
labelled ``b'' are for the value of $J$ at which the new attractor appears
for the positive flip state).  We have
found that the low-$J$ attractors remain even if the radiation field intensity
is increased by a factor 100.
Since the low-$J$ attractors appear even in the absence of thermal flipping,
they clearly are not a thermal trapping effect (Lazarian \& Draine 1999a; see 
\S \ref{sec:intro}).  The radiative torque simply drives the grain to a 
state with low $J$.

When a grain undergoes a thermal flip, $\bJ$ remains fixed in space 
(i.e., $J$ and $\xi$ are unchanged) and $\ahat_1$ flips.  In Figure
\ref{fig:suprathermal_psi70} or \ref{fig:psi70_highJ_noflips}, the point 
representing the grain's state
is reflected in the line $J=0$ when the grain thermally flips.  Thus,
thermal flips do not carry grains across radiative torque-induced 
crossovers.  Because of the near symmetry of the trajectory map, if a 
grain undergoes a thermal flip while approaching one of the low-$J$ 
attractors, it will in most cases land on a trajectory approaching the other 
low-$J$ attractor.  Similarly, a grain that thermally flips while approaching
one of the high-$J$ attractors will usually land on a trajectory approaching
the other high-$J$ attractor.  Due to the presence of the repellors at
$\cos \xi \approx 0.75$, there are some cases where a grain that is 
approaching one of the low-$J$ attractors can be redirected towards one of
the high-$J$ attractors (or vice versa), but $\cos \xi$ must start off quite 
close to $0.75$ for this to happen.  

\begin{figure}
\epsscale{1.00}
\plotone{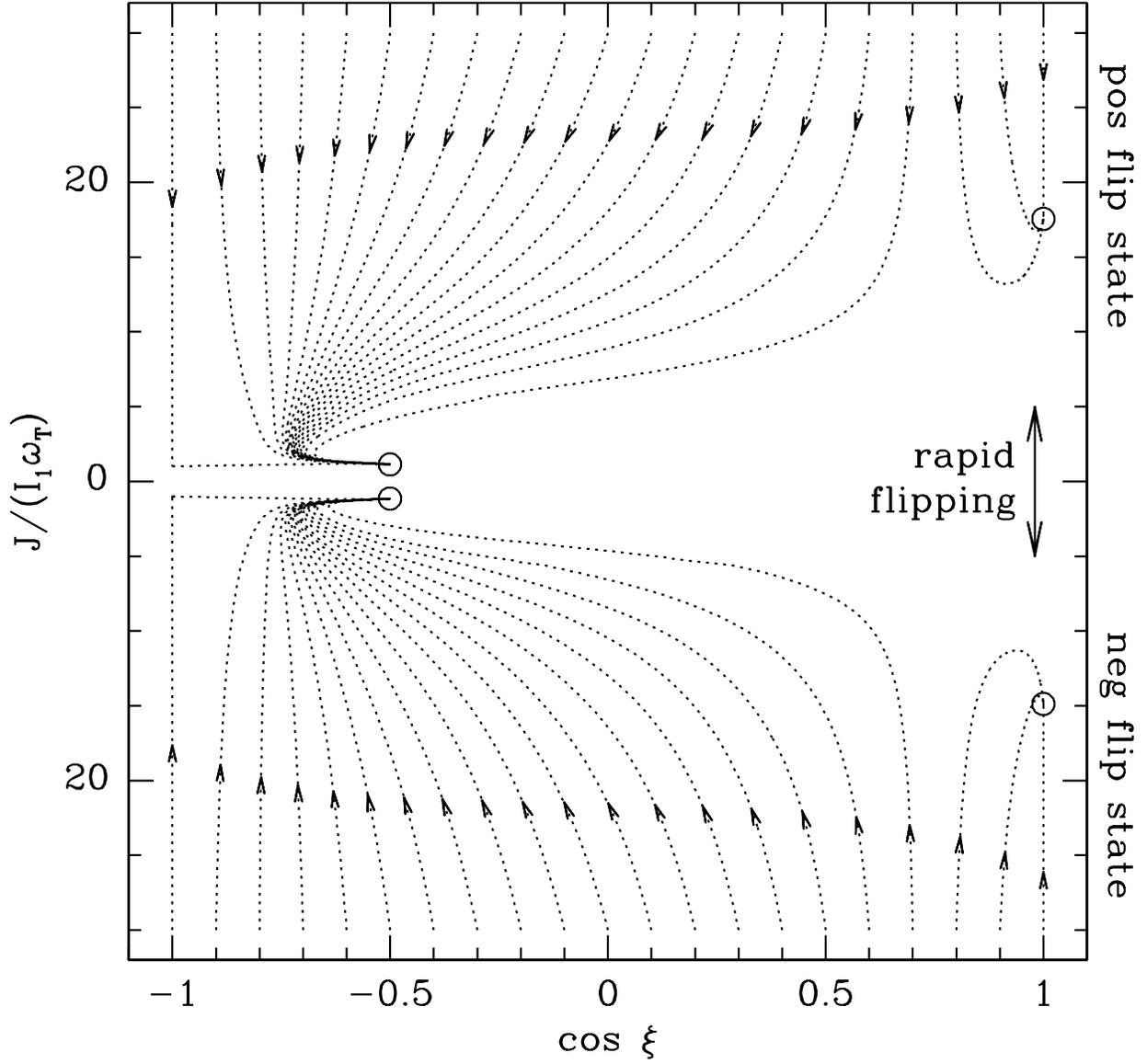}
\caption{
\label{fig:psi70_highJ}
Same as Fig.~\ref{fig:psi70_highJ_noflips}, but with thermal flips partially 
included (see text for details).  The grain 
undergoes rapid flipping when $J/I_1 \omega_{\rm T} \ltsim 5$.  
        }
\end{figure}

Thus, the description of the grain dynamics using deterministic trajectories
is an excellent approximation in this case.  This is not necessarily true
generically, but relies on the high degree of symmetry in the map.  Thermal
flipping does, however, change the appearance of the trajectories.  For 
sufficiently small $J$, the grain undergoes rapid flipping, so the 
trajectory map must be symmetric in reflection about the line $J=0$.  
In Figure \ref{fig:psi70_highJ}, we display trajectories constructed using our
full, stochastic evolution algorithm (for $J/I_1 \omega_{\rm T}=30$ initially).
For trajectories in the upper (lower) half-plane, the grain is initially in
the positive (negative) flip state.  The expectation values $f_+$ and 
$f_-$ are not explicitly indicated at any point along the trjactories, but 
it is clear that $f_+ \approx 1$ when $J/I_1 \omega_{\rm T} \gtsim 
5$ (for trajectories that start in the positive flip state) and 
$f_+ \approx f_- \approx 0.5$ when $J/I_1 \omega_{\rm T} \ltsim 5$.  The 
transition from virtually no flipping to rapid flipping occurs when 
$J/I_1 \omega_{\rm T} \sim 5$, as expected from Figure \ref{fig:timescales2}.
We have displayed trajectories resulting from one iteration of our 
stochastic evolution code; other realizations look nearly identical.

The low-$J$ attractors fall in the rapid-flipping regime; the rapid flipping 
shifts their position to $(J/I_1 \omega_{\rm T},\cos \xi) = (1.14, 
-0.50)$.  Although Figure \ref{fig:psi70_highJ} gives the 
impression that there are two low-$J$ attractors, these are really the same,
because of the rapid flipping.  

\begin{figure}
\epsscale{1.00}
\plotone{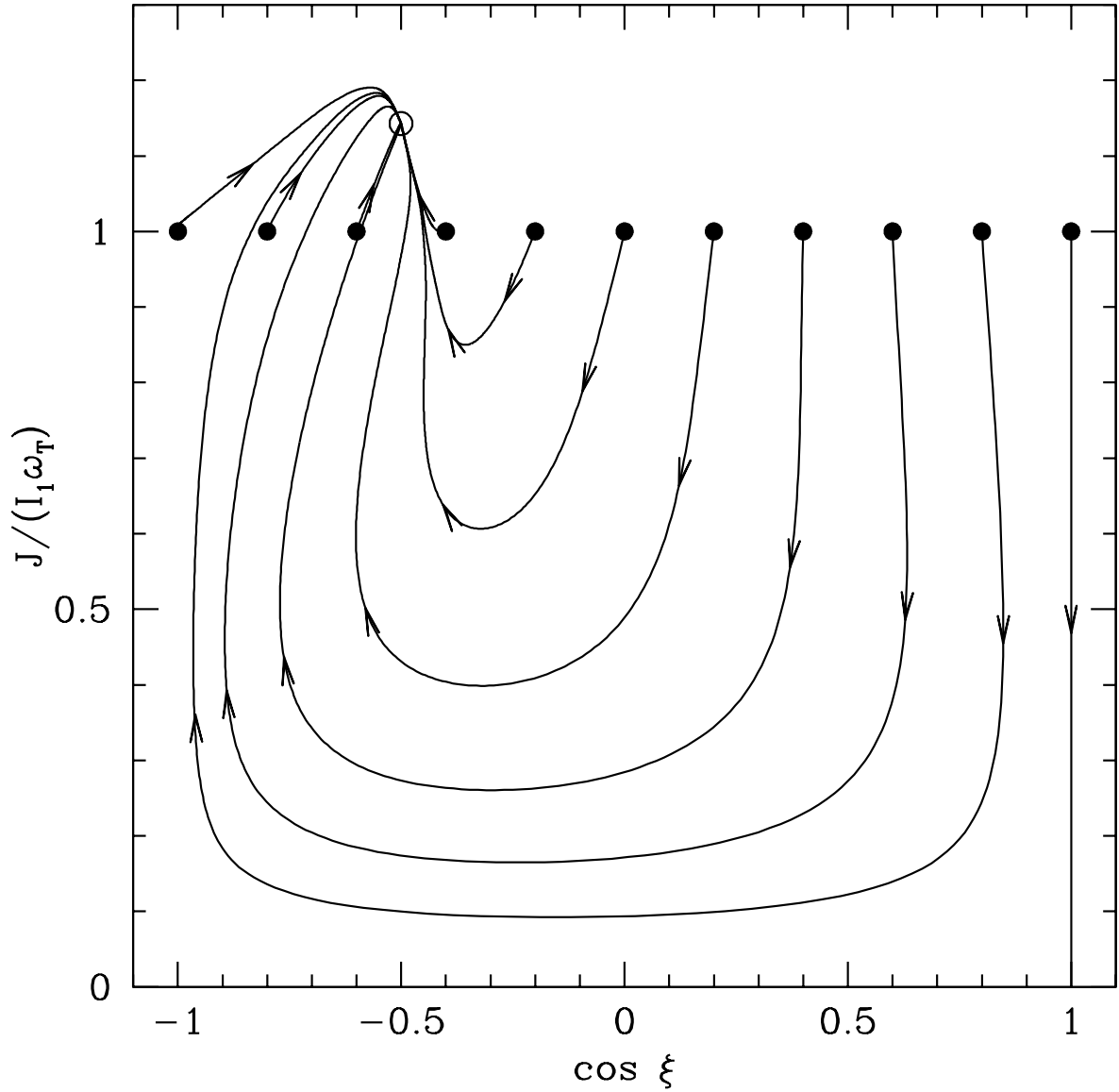}
\caption{
\label{fig:psi70_lowJ}
Same as Figure \ref{fig:psi70_highJ}, but for trajectories starting with 
$J/I_1 \omega_{\rm T}=1$ (initial points are indicated by filled circles).
The attractor, on which most trajectories end, is indicated by an open circle.
The grain undergoes rapid flipping for all points in this figure.  
        }
\end{figure}

In Figure \ref{fig:psi70_lowJ}, we plot the map for trajectories starting
with $J/I_1 \omega_{\rm T}=1$.  The grain is always in the rapid-flipping 
regime for the low values of $J$ in Figure \ref{fig:psi70_lowJ}.
All of these trajectories land on the new, low-$J$
attractor (except for physically irrelevant trajectories with 
$\cos\xi = 1$ initially).  We have found that when
$J/I_1 \omega_{\rm T} < 2$ initially, all trajectories land on the 
low-$J$ crossover.  Thus, 
if a population of real interstellar grains were characterized
by a trajectory map similar to this one, then the high-$J$ 
attractors might not be relevant, since the grains are unlikely to be
in an initial state of suprathermal rotation (although only a very mild
initial suprathermality is needed for some grains to end up at the 
suprathermal attractors).  

We have also analyzed the grain dynamics when $\psi =0$ and $30\arcdeg$.   
The method of Paper II yields a semi-cyclic map with two suprathermal 
attractors when $\psi=0$ and a cyclic map when $\psi=30\arcdeg$.  In both
of these cases, the method developed in this paper yields a non-cyclic map
with a low-$J$ attractor.  

\section{\label{sec:summary} Summary}

In Paper II, we derived equations of motion for a grain exposed to 
radiative torques, under the simplifying assumption that the angular 
momentum $\bJ$ is always parallel to the principal axis of greatest moment 
of inertia $\ahat_1$.  We found that radiative torques can spin grains 
up to suprathermal rotational speeds and (in combination with the 
Barnett torque) align the grains with the interstellar magnetic field. 
In addition, radiative torques can drive grains into periods of thermal
rotation, when they might undergo a type of crossover in which the 
direction of $\bJ$ reverses.  

Because of internal fluctuations, $\ahat_1$ and $\bJ$ are not necessarily 
parallel when $J$ is small.  Thus, we were unable in Paper II to follow the 
grain dynamics during periods of thermal rotation.  Consequently, we were 
unable to ascertain the role of crossovers in the dynamics.  

Here, we have generalized the treatment of Paper II, relaxing the constraint
that $\ahat_1 \parallel \bJ$ and including the effects of thermal 
fluctuations.  Our principal results are:

\noindent 1.  We provide a detailed description of torque-free rotation (\S
\ref{sec:torque-free}), considering the most general grain shape, with no 
degeneracy in the eigenvalues of the inertia tensor.  The rotation is 
completely characterized by the angular momentum $\bJ$, the rotational 
energy $E$, and the ``flip state'' (\S \ref{sec:grain_flips}).  For 
convenience, we define a dimensionless measure of the energy at constant 
$J$, $q \equiv 2 I_1 E/J^2$.  

\noindent 2.  We derive approximate expressions for the external torques,
averaged over torque-free rotation (which occurs on a much shorter 
timescale than the dynamical timescales associated with the torques),
as functions of $\bJ$, $q$, and flip state.  We consider the torques due
to the Barnett magnetic moment, gas and IR emission 
drag, H$_2$ formation, paramagnetic dissipation, and radiation.   

\noindent 3.  We make an order-of-magnitude estimate of the Barnett 
dissipation rate (\S \ref{sec:Barnett_diss}) and find that internal 
dissipation dominates the external torques in determining the evolution 
of $q$.

\noindent 4.  We derive the density of states in the parameter $q$ and find 
expressions for the thermal averages of the torques for various cases of 
interest (\S \ref{sec:q_avg}).  

\noindent 5.  We develop an algorithm for evolving the grain dynamics
(\S \ref{sec:strategy_just_flips}).  The algorithm allows for random thermal 
flips.  

\noindent 6.  As an illustration of the method, we calculate
a trajectory map for one particular case (\S \ref{sec:results}).  Using 
the method of Paper II, we find two attractors (characterized by 
suprathermal rotation) and two crossover points.  The more complete treatment
developed here yields no crossover points and a third attractor, 
characterized by thermal rotation and rapid flipping.  Thus, 
there are qualitative differences introduced by the proper inclusion
of thermal fluctuations and flipping:
whereas the
Paper II analysis produces a semi-cyclic map, the new analysis produces
a non-cyclic map.

In future work, we will investigate more cases, in an effort to determine
whether or not non-cyclic maps with low-$J$ attractors
are a generic feature when the 
analysis developed here is applied.  For this, it will be crucial to 
employ the full ISRF, rather than a monochromatic spectrum with 
$\lambda = 1.2 \micron$.  In addition, we will explore the effect of 
stochastic gas atom impacts and H$_2$ formation events on the dynamics.
Ultimately, we will conduct a systematic study of grain alignment by 
radiative torques, considering a range of shapes and sizes.  

\acknowledgements
This research was supported by NSERC of Canada, NSF grant AST 9988126, and an 
NSF International Research Fellowship to J. C. W.  We are grateful to 
A. Lazarian for many stimulating discussions.

\appendix

\section{Angles in Scattering Coordinates}

When $\ahat_1 \cdot \ehat_3 = 0$, $\Phi = 0$ if $\ahat_1 \cdot
\ehat_2 \ge 0$ and $\Phi = \pi$ if $\ahat_1 \cdot \ehat_2 < 0$.

When $\sin \Theta = 0$, 
\be
\beta = 2 \tan^{-1} \left( \frac{1 - \ahat_2 \cdot \ehat_2 \cos \Theta}
{\ahat_2 \cdot \ehat_3} \right)~~~.
\ee
If $\ahat_2 \cdot \ehat_3 = 0$, then $\beta = \pi - \Theta$ if
$\ahat_2 \cdot \ehat_2 = -1$ and $\beta = \Theta$ if $\ahat_2 \cdot \ehat_2 
= 1$.  

When $\sin \Theta \neq 0$ and $\ahat_2 \cdot \ehat_3 \cos \Phi = \ahat_2 
\cdot \ehat_2 \sin \Phi$, $\beta = 0$ if $\ahat_2 \cdot \ehat_1 <0$
and $\beta = \pi$ if $\ahat_2 \cdot \ehat_1 > 0$.

\section{Dot Products Needed in \S \ref{sec:radtorque_uni}}

\be
\ahat_i \cdot \ehat_1 = \cos\psi \, \zB \cdot \ahat_i - \sin\psi \, \xB \cdot
\ahat_i~~~,
\ee
\be
\ahat_i \cdot \ehat_2 = \cos\psi \, \xB \cdot \ahat_i + \sin\psi \, \zB \cdot
\ahat_i~~~,
\ee
\be
\ahat_i \cdot \ehat_3 = \yB \cdot \ahat_i~~~,
\ee
with 
\be
\xB \cdot \ahat_1 = \cos \xi \cos \phi \sin
\zeta \sin \gamma + \sin \phi \cos \zeta \sin \gamma + \sin \xi \cos \phi
\cos \gamma~~~, 
\ee
\be
\zB \cdot \ahat_1 = 
\cos \xi \cos \gamma - \sin \xi \sin \zeta \sin \gamma~~~, 
\ee
\be
\yB \cdot \ahat_1 = 
\cos \xi \sin \phi \sin \zeta \sin \gamma - \cos
\phi \cos \zeta \sin \gamma + \sin \xi \sin \phi \cos \gamma~~~,
\ee
\be
\xB \cdot \ahat_2 = \cos \xi \cos \phi \left( \cos \alpha \cos \zeta - 
\sin \alpha \sin \zeta \cos \gamma \right) - \sin \phi \left( \cos \alpha
\sin \zeta + \sin \alpha \cos \zeta \cos \gamma \right) + \sin \xi \cos \phi
\sin \alpha \sin \gamma~~~,
\ee
\be
\zB \cdot \ahat_2 = - \sin \xi \left( \cos \alpha \cos \zeta - \sin \alpha
\sin \zeta \cos \gamma \right) + \cos \xi \sin \alpha \sin \gamma~~~,
\ee
and
\be
\yB \cdot \ahat_2 = \cos \xi \sin \phi \left( \cos \alpha \cos \zeta - 
\sin \alpha \sin \zeta \cos \gamma \right) + \cos \phi \left( \cos \alpha
\sin \zeta + \sin \alpha \cos \zeta \cos \gamma \right) + \sin \xi \sin \phi
\sin \alpha \sin \gamma~~~.
\ee

\section{Calculation of $\Fphiavgpm(\xi)$ when $\sin\xi \ll 1$
	\label{app:sinxi->0}}

When $\sin\xi \ll 1$, $\Qq(\xi, \phi) \approx \Qq(\xi = 0 \ {\rm or} \ \pi)$ 
is nearly constant as the grain orbits in $\phi$
and, from the definitions of $F$ and $G$ 
(eqs.~\ref{eq:F} and \ref{eq:G}), we can write
\be
\langle F \rangle_{\pm}(\xi,\phi) = F_{\pm}^0 \cos\xi \cos(\phi-\phi_0) 
- F_{\pm}^1 \sin\xi
\ee
and
\be
\langle G \rangle_{\pm}(\xi,\phi) = -F_{\pm}^0 \sin(\phi-\phi_0)~~~,
\ee
where 
\be
\label{eq:F_bar_q_xi0}
F_{\pm}^0 \equiv 
\left[ \left( \Qq \cdot \ehat_1 \sin \psi
- \Qq \cdot \ehat_2 \cos \psi \right)^2 + 
\left( \Qq \cdot \ehat_3
	\right)^2 \right]^{1/2}~~~,
\ee
\be
F_{\pm}^1 \equiv \Qq \cdot \ehat_1 \cos\psi + \Qq \cdot \ehat_2 \sin\psi~~~,
\label{eq:F_pm^1}
\ee
and the angle $\phi_0$ is determined by
\begin{eqnarray}
\sin\phi_0 &=& \Qq\cdot\ehat_3/F_{\pm}^0
\\
\cos\phi_0 &=& (-\Qq\cdot\ehat_1\sin\psi+\Qq\cdot\ehat_2\cos\psi)/F_{\pm}^0~~~.
\end{eqnarray}
The equation of motion (\ref{eq:eq_of_motion_phi}) for $\phi$ becomes
\be
\frac{d\phi}{dt} = \langle q\rangle \Omega_{\rm B}
\left[1- \frac{g}{\sin\xi} \sin(\phi-\phi_0)\right]~~~,
\ee
where
\be
g \equiv \frac{\gamma_{\rm rad}u_{\rm rad} a^2 \tilde{\lambda}}
	{2 J \langle q\rangle \Omega_{\rm B}} F_{\pm}^0
  \sim 8 \times 10^{-6} \left( \frac{u_{\rm rad}}{u_{\rm ISRF}}
\right) \left( \frac{\tilde{\lambda}}{1.2 \micron}\right) 
\left( \frac{F_{\pm}^0}{10^{-3}}\right) \left( \frac{a}{0.1 \micron}
\right)^{3/2} \left( \frac{\omega}{\omega_{\rm T}} \right)^{-1}~~~;
\label{eq:gdef}
\ee
in equation (\ref{eq:gdef}), we have taken $\langle q \rangle \approx 1$, 
$\gamma_{\rm rad} = 0.1$, $T_{\rm gas} = 100 \K$, 
$\chi_0 = 3.3 \times 10^{-4}$, and $B = 5 \, \mu {\rm G}$.  

A transition from periodic motion in $\phi$ to no motion in 
$\phi$ occurs when $\sin\xi=g$.  When $\sin\xi > g$,
the motion in $\phi$ is periodic with period
\be
\tau_\phi = \frac{2\pi}{\langle q\rangle \Omega_{\rm B}} 
\left[1-\left(\frac{g}{\sin\xi}\right)^2\right]^{-1/2}
\ee
and $\cos(\phi - \phi_0)$ averages to zero over the motion in $\phi$.  
When $\sin\xi\le g$, the dynamics is characterized by a stable stationary 
point $\phi_s$ given by
\begin{eqnarray}
\sin(\phi_s-\phi_0) &=& \sin\xi/g \\
\cos(\phi_s-\phi_0) &=& \left[1-(\sin\xi/g)^2\right]^{1/2}
\end{eqnarray}
Thus, 
\be
\label{eq:Fphiavgpm_small_sin_xi}
\Fphiavgpm(\xi) = \cases{-F_{\pm}^1 \sin\xi &, $g \leq \sin\xi \ll 1$\cr
-F_{\pm}^1 \sin\xi + F_{\pm}^0 \cos\xi \left[1-(\sin\xi/g)^2\right]^{1/2} &, 
$\sin \xi \le g$, ~~$\sin\xi\ll 1$\cr}~~~.
\ee

As discussed in \S \ref{sec:strategy_just_flips}, we construct
interpolation tables in $\cos\xi$ and $J^2/(2I_1 kT_d)$ for $\Fphiavgpm$ and 
$\Hphiavgpm$, to be used in evolving the grain dynamics.  However, we 
use equation (\ref{eq:Fphiavgpm_small_sin_xi}) to directly evaluate 
$\Fphiavgpm$ when $\xi$ lies within one interpolation zone of 0 or $\pi$.
Otherwise, it would be impossible to reproduce the small-$\sin\xi$ 
structure with a reasonable number of interpolation zones.  Thus, we also
construct a table of $F_{\pm}^0[J^2/(2I_1 kT_d)]$.  To ensure continuity
at $\xi_1 =$ the first interpolation point away from $\xi=0$ or $\pi$, we
estimate $F_{\pm}^1 \approx - \Fphiavgpm(\xi_1) / \sin\xi_1$.  

\section{Derivation of $P_{\rm same}$}

The probability that exactly $n$ flips occur during $\Delta t$ is
\be
P_n = \frac{1}{n!} \kappa^n e^{- \kappa}~~~,
\ee
where $\kappa \equiv \Delta t / \tau_{\rm tf}$.  The probability of zero
flips or an even number of flips is
\begin{eqnarray}
P_{\rm same} = \sum_{j=0}^{\infty} P_{2j} &=& e^{-\kappa} \sum_{j=0}^{\infty}
\frac{\kappa^{2j}}{(2j)!} \nonumber \\
&=& e^{-\kappa} \frac{1}{2} \left[ \sum_{j=0}^{\infty} \frac{\kappa^j}{j!} + 
\sum_{j=0}^{\infty} \frac{(-\kappa)^j}{j!} \right] \nonumber \\
&=& e^{-\kappa} \frac{1}{2} \left( e^{\kappa} + e^{-\kappa} \right) 
= e^{-\kappa} \cosh(\kappa)~~~.
\end{eqnarray}

\newpage
\section{Glossary of Notation}
\begin{table}[h]
\begin{center}
\begin{tabular}{c l}
symbol	& significance\\
\hline
$a$	& radius of sphere of equal volume, $a=(3V/4\pi)^{1/3}$\\
$\ahat_1$, $\ahat_2$, $\ahat_3$ & grain principal axes (eigenvectors of moment of inertia tensor)\\
$I_1,I_2,I_3$	& eigenvalues of moment of inertia tensor ($I_1\ge I_2 \ge I_3$)\\
$\bJ$	& angular momentum of grain\\
$\omega_{\rm T}$ & rotation rate for a sphere in thermal equilibrium with 
	           the gas (eq.~\ref{eq:omega_T})\\ 
$q$	& $E/(J^2/2I_1)$ (eq.~\ref{eq:q_def})\\
$T_d$	& vibrational temperature of grain\\

$\tau_{\rm Bar}$ & at fixed $J$, timescale for Barnett damping \\
	&of rotational kinetic energy in excess of $J^2/2I_1$
	(eq.~\ref{eq:tau_Bar})\\
$\bar{A}(q,\pm)$ & time-average of dynamical variable $A$ over torque-free motion \\
	& with $E/(J^2/2I_1)=q$ and in $\pm$ flip state
	(eq.~\ref{eq:bar_A})\\
$\tilde A$ & average over radiation field spectrum 
	(eqs.~\ref{eq:tilde_lambda}, \ref{eq:tilde_Q})\\
$\langle A\rangle_\pm(J,T_d,\xi,\phi)$ &
	thermal average of dynamical variable $A$\\
	& for grain with given $\bJ$ and $T_d$, in $\pm$ flip state 
	with respect to $\ahat_1$
	(eq.~\ref{eq:thermal_avg_q})\\
$\langle A\rangle_\pm^\phi(J,T_d,\xi)$ &
	thermal average of variable $A$ for grain
	with given $J$ and $T_d$, 
	in $\pm$ flip\\
	& state with respect to $\ahat_1$,
	averaged over motion in $\phi$ (eqs.~\ref{eq:Fphiavgpm}, 
	\ref{eq:Hphiavgpm})\\
$\tau_{\rm tf}^{-1}$ & probability per unit time that grain will change its
	flip state (eq.~\ref{eq:tau_tf})\\
$P_0(\Delta t)$	& probability that grain undergoes no flips during interval 
	$\Delta t$\\
$P_{\rm same}(\Delta t)$ & probability that flip state
	at end of interval \\
	&will be same as flip state at beginning of
	interval (eq.~\ref{eq:P_same})\\
$f_{\rm same}(\Delta t)$ & 
	expectation value for fraction of time that grain \\
	&will spend in initial flip state during interval $\Delta t$
	(eq.~\ref{eq:f_same})\\
$f_{\rm s}(\Delta t)$ & expectation value for fraction of time spent in 
	original flip state\\
	& for grain which flips at least once during interval
	$\Delta t$ (eq.~\ref{eq:f_s})\\
$\bQ_\bGam(\Theta,\Phi,\beta,\lambda)$ & radiative torque efficiency vector
	for grain orientation $(\Theta,\Phi,\beta)$ \\
	&relative to
	direction of incident radiation with wavelength $\lambda$
	(eq.~\ref{eq:Gamma_rad})\\
\end{tabular}
\end{center}
\end{table}


\begin{deluxetable}{lll}
\tablecaption{Branching Probabilities\label{tab:branches}}
\tablehead{
\colhead{probability}&
\colhead{$f_0$}&
\colhead{final flip state}
}
\startdata
$P_0$		& 1		& original (no flips)\\
$(1-P_0)(1-\Pf)$	& $f_s$		& original (even \# of flips)\\
$(1-P_0)\Pf$	& $f_s$		& opposite (odd \# of flips)\\
\enddata
\end{deluxetable}

\end{document}